\documentclass[11pt,a4paper]{article}
\usepackage{jcappubnew}

\newcommand{\beq}{\begin{equation}}
\newcommand{\eeq}{\end{equation}}
\newcommand{\bea}{\begin{eqnarray}}
\newcommand{\eea}{\end{eqnarray}}

\newcommand{\rmi}{\mathrm{i}}

\newcommand{\vc}[1]{{\boldsymbol #1}}

\newcommand{\lh}{\left(}
\newcommand{\rh}{\right)}
\newcommand{\der}{\partial}
\renewcommand{\d}{\mathrm{d}}
\newcommand{\non}{\nonumber}
\DeclareMathSymbol{\mg}{\mathrel}{symbols}{"1D}

\newcommand{\get}{\eta}

\newcommand{\gx}{\xi}

\newcommand{\gs}{\sigma}

\newcommand{\gf}{\phi}

\newcommand{\gc}{\chi}

\newcommand{\f}{f_\mathrm{NL}}

\newcommand{\tg}{{\tilde g}}

\newcommand{\tn}{{\tilde n}}

\newcommand{\bv}{{\bar v}}

\newcommand{\bdm}{\begin{displaymath}}
\newcommand{\edm}{\end{displaymath}}
\newcommand{\nn}{\nonumber}
\def\be{\begin{equation}}
\def\ee{\end{equation}}

\def\e{\epsilon}

\def\hpa{\eta^{\parallel}}
\def\hpe{\eta^{\perp}}

\begin{document}

\title{Momentum dependence of the bispectrum in two-field inflation}

\author{Eleftheria Tzavara}
\author{and Bartjan van Tent}
\affiliation{Laboratoire de Physique Th\'eorique, Universit\'e Paris-Sud 11 
and CNRS, B\^atiment 210, 91405 Orsay Cedex, France}
\emailAdd{Eleftheria.Tzavara@th.u-psud.fr} 
\emailAdd{Bartjan.Van-Tent@th.u-psud.fr}

\abstract{We examine the momentum dependence of the bispectrum of two-field 
inflationary models within the long-wavelength formalism. We determine 
the sources of scale dependence in the expression for the parameter of 
non-Gaussianity $\f$ and study two types of variation of the momentum 
triangle: changing its size and changing its shape. We introduce two 
spectral indices that quantify the possible types of momentum dependence 
of the local type $\f$ and illustrate our results with examples.}
 \begin{flushright}
 LPT-12-116 \\ 
 \end{flushright}

\maketitle
\flushbottom

\section{Introduction}\label{intro}

The study of inflationary non-Gaussianities and their impact on 
the cosmic microwave background has been an important 
subject of cosmological research in recent years. In nine years of WMAP data 
\cite{Bennett:2012fp} and one year of data from the Planck satellite 
\cite{Ade:2013ydc} no primordial non-Gaussianity of the local and equilateral
types (see below) was observed, and constraints have tightened considerably.
Next year's Planck release is expected to put even tighter constraints on 
those types of non-Gaussianity, as well as investigate many additional types
(which differ in their momentum dependence).
The importance of the current constraints and a future possible detection
or further improvements of the constraints lies in the fact that they allow
us to discriminate between different classes of models of inflation, since these
predict different types and amounts of non-Gaussianity.

There are basically two distinct types of non-Gaussianity that are most important from the point of view of inflation: the equilateral type 
produced at horizon-crossing, which has a quantum origin and is  
maximal for equilateral triangle configurations \cite{Creminelli:2005hu}, 
and the local type produced outside the inflationary horizon due to the existence of interacting fields. The latter is maximized for squeezed triangles, 
i.e.\  isosceles triangles with one side much smaller than the other two \cite{Komatsu:2001rj,Babich:2004yc}. 
The first type is known to be slow-roll suppressed for single-field models with standard kinetic terms and trivial field metric 
\cite{Maldacena:2002vr}. 
On the other hand, some models with non-standard kinetic terms coming from higher-dimensional cosmological models are known to produce non-Gaussianity 
of the equilateral type so large that it is not compatible with WMAP and Planck observations 
\cite{Alishahiha:2004eh,Silverstein:2003hf,Mizuno:2009cv,Mizuno:2010ag}, thus leading people to consider an extra field 
in order to achieve smaller values of the parameter $\f$ of non-Gaussianity. 

Non-Gaussianity of the squeezed type can be found naturally in multiple-field models of inflation \cite{Rigopoulos:2005us,Bernardeau:2002jy}, 
due to the sourcing of the adiabatic mode by the isocurvature components outside the horizon. 
For single-field models this is obviously impossible due 
to the absence of isocurvature modes. There has been much study of two-field models 
\cite{Seery:2005gb,Kim:2006te,Battefeld:2006sz,Battefeld:2007en,
Langlois:2008vk,Cogollo:2008bi,Vernizzi:2006ve,Tzavara:2010ge}, being the easiest to investigate, in the hope of 
finding a field potential that can produce local non-Gaussianity large enough to be measurable in the near-future. It proves to be non-trivial to 
sustain the large non-Gaussianity produced during the turn of the fields until the end of inflation.

Non-Gaussianity produced at horizon-crossing is known to be momentum-dependent. 
The scale dependence of the equilateral $\f$ produced for example from DBI inflation \cite{Langlois:2008qf,Arroja:2008yy,Mizuno:2009cv,Cai:2009hw,Senatore:2010wk}, has been examined both theoretically 
\cite{Chen:2005fe,Khoury:2008wj,Byrnes:2009qy,Leblond:2008gg} and in terms of observational forecasts \cite{LoVerde:2007ri,Sefusatti:2009xu}. 
In this paper we are going to study the scale dependence of local-type models that has not been studied as much. 
Squeezed-type non-Gaussianity, produced outside the horizon, is usually associated with a parameter of non-Gaussianity 
$\f^\mathrm{local}$ that is local in real space, and therefore free of any explicit momentum dependence, defined through 
$\zeta(x)=\zeta_{L}(x)+(3/5)\f^\mathrm{local}(\zeta_{L}(x)^2-\langle\zeta_L(x)\rangle^2)$, where $\zeta_{L}$ is the linear Gaussian part. 
Nevertheless, calculations of $\f$ for several types of multiple-field models (see e.g.\  \cite{Vernizzi:2006ve,Byrnes:2008wi,
Tzavara:2010ge}) show that there is always a momentum dependence inherited from the horizon-crossing era, which can in principle result in 
a tilt of $\f$.  
When a physical quantity exhibits such a tilt one usually introduces a spectral index, as for example in the case of the power spectrum.  
The observational prospects of the detection of this type of scale dependence of local $\f$ were studied in \cite{Sefusatti:2009xu}.  
Only recently spectral indices for $\f$ were defined in \cite{Byrnes:2010ft,Byrnes:2009pe,Byrnes:2012sc}, keeping constant the shape of the 
triangle or two of its sides, within the 
$\delta N$ formalism.  
Note, however, that most theoretical predictions have considered equilateral triangles for simplicity, even though the local-type 
configuration is maximal on squeezed triangles. 
If one were to calculate a really squeezed triangle, then $\f^\mathrm{local}$ acquires some intrinsic momentum dependence 
due to the different relevant scales, as was shown in \cite{Tzavara:2010ge}. 

It is both these effects we want to study in this paper: 
on the one hand the tilt of $\f$ due to the background evolution at horizon-crossing and on the other hand the impact of the shape of the triangle on $\f$.  
In order to do that in a concrete way, such that these effects do not mix, we define two independent spectral indices, each one quantifying different deformations of the momentum triangle. 
Moreover, having an exact expression of $\f$ for an isosceles triangle, we are able to study and understand for the first time the origin of both types 
of momentum dependence of $\f$. We also provide analytical estimates for the quadratic model (which actually hold for any equal-power sum model) that we use in this paper to illustrate our results. 

The paper is organised as follows. We begin by 
reviewing the power spectrum and the bispectrum along with the 
background of the theory in section \ref{pre}. 
In section \ref{sour} 
we present the long-wavelength formalism results and discuss the sources of scale 
dependence in the expression for $\f$. We also introduce two  
spectral indices, able to quantify the effects of different triangle 
deformations. 
In section \ref{conf} we study the scale dependence for triangles of constant shape but of varying size, which is mainly due
to horizon-crossing quantities, while 
in section \ref{shape} we study the scale dependence related to the shape of 
the triangle. Finally, we conclude in section \ref{concl}, while analytical 
expressions for the final value of 
the spectral indices for any equal-power sum potential are given in Appendix \ref{app}.

\section{Preliminaries}\label{pre}

The power spectrum of the cosmological adiabatic 
perturbation $\zeta_1$  
has proved to be a valuable tool to connect the theory of inflation to 
observations of the sky. The power spectrum provides two important 
observables: its amplitude $\mathcal{P}_{\zeta}(k,t)$,
\be
\langle\zeta_{1\vc{k_1}}\zeta_{1\vc{k_2}}\rangle
 = \delta^3(\vc{k_1}+\vc{k_2})\frac{2\pi^2}{k_1^{3}}\mathcal{P}_{\zeta}
(k_1),
\ee
carrying information about the mass scale of the inflaton,  
and the spectral index $n_{\zeta}$, quantifying the tilt of this almost 
constant, scale-independent spectrum due to the small (assuming slow roll) but non-zero 
evolution of the inflationary background,
\be
n_{\zeta} - 1 \equiv\frac{\d\ln{\mathcal{P}_{\zeta}}}{\d\ln{k}}
= \frac{\d\ln{\mathcal{P}_{\zeta}}}{\d t_k}\frac{\d t_k}{\d\ln{k}}
=\frac{\d\ln{\mathcal{P}_{\zeta}}}{\d t_k}\frac{1}{1-\e_k}. 
\label{spectral}
\ee
From now on a $k$ subscript on quantities will denote 
evaluation at the time $t_k$ that the scale $k$ exits the horizon. 
The last equality in (\ref{spectral}) comes from the relation 
$k=aH$ at horizon crossing, while using 
\bea
&&\dot{H}=-\e H,\nn\\
&&\dot{a}=a.
\eea
Here we choose the time coordinate to be the number of e-foldings (for details see \cite{Tzavara:2010ge}). The first of the above equations is just 
the definition of the $\e$ slow-roll parameter, while the second is the form that  the definition of the Hubble parameter acquires due to the 
choice of the time variable. 

We also give here the Einstein and field equations \cite{Rigopoulos:2005xx} for the inflaton fields $\phi^A$, with 
the index $A$ numbering the fields, that roll down a potential $W$: 
\bea
&&H^2 = \frac{\kappa^2}{3}\left(\frac{\Pi^2}{2}+W\right),
\qquad\qquad
\dot{H}  =-\frac{\kappa^2 \Pi^2}{2H}, \non\\
&&\dot{\Pi}^A  = -3\Pi^A-\frac{W^{,A}}{H},
\label{fieldeq}
\eea
with $\kappa^2\equiv8\pi G=8\pi/m_{pl}^2$ and $W_{,A}\equiv\der W/\der\gf^A$. $\Pi^A=H\dot{\phi}^A$ is the canonical momentum of the inflaton. 
In this paper we choose to work with two scalar 
fields, so that $A=1,2$, with canonical kinetic terms and a trivial field metric. Generalizing  
any of these assumptions would still allow for the system to 
be studied in principle, although the computations would be more complicated.

Apart from $\e$, a hierarchy of slow-roll parameters can be built from the time derivatives of the inflation fields:
\be
\eta^{(n)A}\equiv\frac{1}{H^{n-1}\Pi}\lh H\frac{\d}{\d t}\rh^{n-1}\Pi^A.
\ee
We will denote the first and second-order slow-roll parameters of the 
hierarchy as $\eta^{(1)A}\equiv\eta^A$ and $\eta^{(2)A}\equiv\xi^A$, respectively. 
In order to distinguish adiabatic from isocurvature effects we construct 
a basis in field space $\{e_1^A,e_2^A\}$, where $e_1^A$ is a unit vector 
parallel to the velocity 
of the fields and $e_2^A$ a unit vector parallel to the component of the 
acceleration that is perpendicular to the velocity:
\be
e_1^A=\frac{\Pi^A}{\Pi}\qquad\mathrm{and}\qquad
e_2^A=\frac{\dot{\Pi}^A-e_1^Ae_{1B}\dot{\Pi}^B}
{|\dot{\Pi}^A-e_1^Ae_{1B}\dot{\Pi}^B|}.
\ee 
This is why we use the $1$ subscript for the adiabatic perturbation $\zeta_1$, while the 
isocurvature perturbation will be denoted as $\zeta_2$. 
The projection of the slow-roll parameters on $e_1^A$ will be denoted 
by $\parallel$ and the projection on $e_2^A$ by $\perp$. Finally 
we also define the slow-roll parameter $\chi$ as
\be
\chi=\frac{W_{22}}{3H^2}+\e+\hpa,
\ee
where $W_{22}=W^{,AB}e_{2A}e_{2B}$ is the second derivative of the 
potential $W$ of the fields projected in the $22$ direction.

In addition to the power spectrum we can gain more information from the CMB 
by studying the Fourier transform of the three-point correlation function, 
\bea
\langle \zeta_{1\vc{k_1}} \zeta_{1\vc{k_2}} \zeta_{1\vc{k_3}}\rangle
& \equiv & (2\pi)^{-3/2}\delta^3 (\sum_s\vc{k_s})B_{\zeta}(k_1,k_2,k_3),
\eea
where $B_{\zeta}$ is the bispectrum. Because of the overall $\delta$-function we see that the vectorial sum of the 
three $k$-vectors has to be zero. In other words, the three $k$-vectors form a triangle. 
The amplitude of the bispectrum can 
provide additional constraints on the slow-roll parameters of a given type 
of inflationary model. The profile of the bispectrum, i.e.\  the shape of the momentum triangle, 
gives information on the type of the 
inflationary model itself. For example, models with higher-order kinetic 
terms produce a bispectrum of the equilateral type (see e.g.~\cite{Komatsu:2010hc}), mainly due to quantum 
interactions at horizon crossing. By equilateral type we mean a 
bispectrum that becomes maximal for equilateral triangles. On the other hand,  
canonical multiple-field inflation models predict a bispectrum of the local type. 
This arises from non-linearities of the form $\zeta_1=\zeta_{1L}-(3/5)\f(\zeta_
{1L}^2-\langle\zeta_{1L}\rangle^2)$ ($\zeta_{1L}$ being the first-order adiabatic perturbation) that 
are created classically outside the horizon, leading to a bispectrum of the form
\be
B_{\zeta}(k_1,k_2,k_3)=-\frac{6}{5}\f\lh \frac{2\pi^2}{k_1^3}
\mathcal{P}_{\zeta}(k_1)\frac{2\pi^2}{k_2^3}\mathcal{P}_{\zeta}(k_2)
+\lh k_2\leftrightarrow k_3\rh
+\lh k_1\leftrightarrow k_3\rh\rh,\label{bisp}
\ee
where $\f$ is usually assumed to be constant.
This bispectrum becomes maximal for a squeezed triangle, i.e.\  
a triangle with two sides almost equal and much larger than the 
third one. 
As we will discuss in the rest of the paper, $\f$ is not actually a constant,
but depends on the size and shape of the momentum triangle.

\begin{figure}
\begin{center}
\includegraphics[scale=0.2]{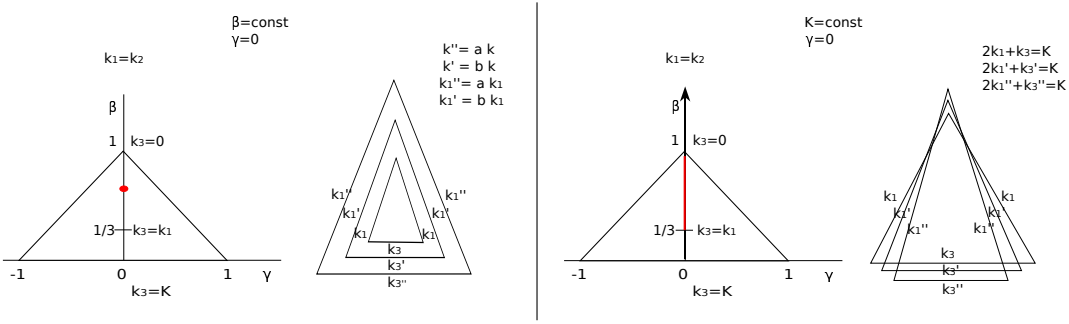}
\end{center}
\caption{The types of deformations of the momentum triangle we are 
considering. Left: Conformal transformation of the 
triangle. In the $\gamma,\beta$ plane this corresponds to a 
constant (red) point. Right: Keeping the perimeter of the triangle 
$K$ constant we change the shape of an isosceles 
$\gamma=0$ triangle, moving along the bold red line for 
$k_1=k_2\geq k_3$.}
\label{fig01}
\end{figure}

In order to study the dependence of the non-Gaussianity on the 
shape of the triangle, instead 
of using $k_1,k_2$, and $k_3$ we will use the variables introduced in 
\cite{Rigopoulos:2004ba,Fergusson:2008ra}, 
\bea
K=\frac{k_1+k_2+k_3}{2},\qquad\gamma=\frac{k_1-k_2}{K},
\qquad\beta=-\frac{k_3-k_1-k_2}{2K},\label{newvar}
\eea
which correspond to the perimeter of the triangle and two scale ratios 
describing effectively the angles of the triangle. They have the following 
domains: $0\leq K\leq \infty,\ 0\leq\beta\leq1$ and 
$-(1-\beta)\leq\gamma\leq1-\beta$, see figure~\ref{fig01}. 
As one can check from 
the above equations,  
the local bispectrum becomes maximal for $\beta=1$ and $\gamma=0$,   
or $\beta=0$ and $\gamma=\pm 1$, i.e.\  for a squeezed triangle. 
In this paper we always assume 
$k_1=k_2$, dealing only with equilateral or isosceles triangles
(note that the relation $k_1=k_2$ is satisfied by definition for both 
equilateral and squeezed triangles).  
The two scales of the triangle $k_3 \equiv k \le k' \equiv k_1 = k_2$ can be expressed 
in terms of the new parameters $\beta$ and $K$ as 
\be
k=(1-\beta)K
\qquad\mathrm{and}
\qquad k'=\frac{1+\beta}{2}K,\label{par}
\ee
while $\gamma=0$. The condition $k\le k'$ means that we only have to study acute isosceles triangles $1/3\leq\beta\leq1$.

\section{Sources of scale dependence}\label{sour}

\subsection{Long-wavelength results}

In this paper we use the long-wavelength formalism to study the 
parameter of 
non-Gaussianity $\f$ and its scale dependence. 
The long-wavelength formalism consists of doing a perturbative expansion
of the exact differential equations for the non-linear cosmological 
perturbations on super-horizon scales, which were obtained by neglecting
second-order spatial gradient terms
(see \cite{Rigopoulos:2005xx,Rigopoulos:2005us,Tzavara:2010ge}).
Slow-roll solutions for the first-order perturbations at horizon-crossing
are used as initial conditions.\footnote{The non-Gaussianity produced 
at horizon-crossing is not included in the long-wavelength formalism, but can 
be computed in another way, see \cite{Tzavara:2011hn}, and 
added by hand. It is negligibly small in models with standard kinetic terms.}
The super-horizon 
assumption is equivalent to the leading order of the spatial gradient expansion and 
requires the slow-roll assumption to be satisfied at horizon-crossing (but not 
afterwards, see \cite{Leach:2001zf,
Takamizu:2010xy,Tzavara:2010ge}). The non-Gaussianity parameter for an isosceles triangle of 
the form $k_1=k_2\equiv k'\geq k_3\equiv k$ was found in \cite{Tzavara:2010ge} to be
\bea
	-\frac{6}{5}\f=\frac{-2\bv_{12k'}/[1+(\bv_{12k'})^2]}
{1+(\bv_{12k'})^2+2\frac{\gamma_k^2}{\gamma_k'^2}[1+(\bv_{12k})^2]}
	\!&\!\!\Bigg[&\!\!\bv_{12k'}\!\Bigg
(\!g_{sr}
	(k',k')\!
	+\!g_{iso}(k',k')\!+\!g_{int}(k',k')\!\Bigg) \non\\	
&&\hspace{-1.5cm}+2\frac{\gamma_k^2}{\gamma_k'^2}\bv_{12k}\Bigg(g_{sr}(k',k)\!+\!g_{iso}(k',k)
\!+\!g_{int}(k',k)\!\Bigg)\Bigg],
\label{fNLgeni}
\eea
where $\f=\f(t;t_{k'},t_k)$ depends on $t_{k'}$ and $t_k$, denoting 
the horizon-crossing times of the two scales $k'$ and $k$ of the triangle,
respectively. 
This result is exact and valid beyond the slow-roll approximation after horizon-crossing. 
All quantities appearing in this formula will be explained below.

The quantity $\bv_{12}$ is a 
transfer function showing how the isocurvature mode (denoted by the subscript $2$) sources  
the adiabatic component $\zeta_1$. 
In the following two more transfer 
functions will appear, namely $\bv_{22}$ and $\bv_{32}$, showing how the isocurvature mode sources the isocurvature 
component $\zeta_2$ and the velocity of the isocurvature component 
$\theta_2 \equiv \dot{\zeta_2}$, respectively. $\bv_{a2}$ is a function of the 
horizon-exit time $t_k$ of the relevant perturbation of scale $k$ and it 
also evolves with time $t$, at least during inflation. In (\ref{fNLgeni}) as 
well as in the formulas that follow, $\bv_{a2k}\equiv\bv_{a2}(t,t_k)$.  
The indices $a,b$ take the values $1,2,3$, indicating respectively the 
adiabatic perturbation $\zeta_1$, the isocurvature perturbation $\zeta_2$,
and the isocurvature velocity $\theta_2$.\footnote{Due to the exact relation
$\theta_1 = 2 \eta^\perp \zeta_2$, there is no need to consider the velocity
of the adiabatic perturbation $\theta_1$ as an additional variable
\cite{Rigopoulos:2005us}.} 
$\bv_{a2}$ comes from the 
combination of the Green's functions $G_{a2}$ and $G_{a3}$ of the system of equations for the 
super-horizon perturbations 
(for the system of equations that 
the Green's functions obey see \cite{Rigopoulos:2005us,Tzavara:2010ge}):
\be
\bv_{a2}(t,t_k) = G_{a2}(t,t_k)-\chi_k G_{a3}(t,t_k). 
\label{green}
\ee
The quantity $\gamma_k$ in (\ref{fNLgeni}) is defined as 
$\gamma_k\equiv-\kappa H_k/(2k^{3/2}\sqrt{\e_k})$ (and is not related to the $\gamma$ defined in (\ref{newvar})).

Except for the overall factor, $\f$ has been split into three 
contributions: $g_{sr}$, $g_{iso}$ and $g_{int}$.\footnote{In 
\cite{Tzavara:2010ge} we had also a fourth contribution $g_k$, denoting the terms that vanish for an equilateral triangle. 
Here we have incorporated these terms in $g_{sr}$ (the last two lines), since they are also slow-roll suppressed.}
$g_{sr}$ is a term that is slow-roll suppressed, since it depends only on horizon-exit quantities, where by assumption slow-roll holds,
\bea
g_{sr}({k}_1,{k}_2)&=&\hpe_{k_1}\Bigg(\frac{G_{22k_1k_2}\bv_{12k_1}}{2}   
-\frac{1}{\bv_{12k_2}}-\frac{G_{22k_1k_2}}{2\bv_{12k_1}} 
\Bigg)+\frac{3\chi_{k_2}}{4}G_{33k_1k_2}
-\frac{3}{2}(\e_{k_1}+\hpa_{k_1})G_{22k_1k_2}\nn\\
&&
+\frac{\chi_{k_1}}{4}\Bigg(2\frac{\bv_{12k_1}}{\bv_{12k_2}}+
G_{22k_1k_2}\Bigg)-\frac{\e_{k_1}+\hpa_{k_1}}{2(\widetilde{v}_{12})^2}
\nn\\
&&
+\frac{G_{13}(t,t_{k_1})}{2}\Bigg[\frac{3\lh\chi_{k_1}G_{22k_1k_2}
-\chi_{k_2}G_{33k_1k_2}\rh}{2\bv_{12k_1}}
+G_{32k_1k_2}\lh\frac{3+\e_{k_1}+2\hpa_{k_1}}{2\bv_{12k_1}}
+\hpe_{k_1}\rh\Bigg]\nn\\
&&
-\frac{3}{4}G_{32k_1k_2}
-\frac{1}{2}G_{12k_1k_2}\lh
\e_{k_1}+\hpa_{k_1}+2\hpe_{k_1}-\frac{\chi_{k_1}}{2}\lh1+\bv_{12k_1}\rh
+\frac{\e_{k_1}+\hpa_{k_1}}{\bv_{12k_1}}\rh.
\eea
Here we introduce some new notation,
\begin{displaymath}
\begin{array}{l}
(\widetilde{v}_{12})^2\equiv\bv_{12k_1}\bv_{12k_2},\qquad 
(\widetilde{v}_{22})^2\equiv\bv_{22k_1}\bv_{22k_2},\qquad
(\widetilde{v}_{32})^2\equiv\bv_{32k_1}\bv_{32k_2},\\  
\widetilde{v}_{22}\widetilde{v}_{32}\equiv\frac{1}{2}(\bv_{22k_1}\bv_{32k_2}
+\bv_{22k_2}\bv_{32k_1}),
\end{array}
\end{displaymath}
and also $G_{abk_1k_2}\equiv G_{ab}(t_{k_1},t_{k_2})$. Moreover, we assume
$k_1 \geq k_2$.
$g_{sr}$ is the only term from which a (small) part survives in the 
single-field limit, i.e.\  in the 
limit where $\bv_{12}=0$ at all times. 
For the equilateral case $k'=k$ the two last lines of $g_{sr}$ are zero, 
since the Green's functions satisfy
\footnote{In addition, in the equilateral case $k'=k$ the $\gamma_k$ ratios 
in (\ref{fNLgeni}) reduce to $1$ and the two terms in the brackets of (\ref{fNLgeni}) become 
identical (apart from the factor $2$).} 
\be
G_{ab}(t,t)=\delta_{ab}.
\label{Green_norm}
\ee

The contribution $g_{iso}$ is a term that survives as long as the isocurvature modes are alive,
\be
g_{iso}({k}_1,{k}_2)=(\e+\hpa)(\widetilde{v}_{22})^2
+\widetilde{v}_{22}\widetilde{v}_{32}.\label{giso}
\ee
If at the end of inflation these are non-zero, $\f$ can still evolve 
afterwards and we cannot be sure that its value survives until today. Finally, 
$g_{int}$ is given by
\be
g_{int}({k}_1,{k}_2)=\!-\!\int_{t_{k_1}}^t\!\!\!\!\d t'\Big[2(\hpe)^2
(\widetilde{v}_{22})^2
\!+\!(\e+\hpa)\widetilde{v}_{22}\widetilde{v}_{32}\!+\!(\widetilde{v}_{32})^2
\!-\!G_{13}(t,t')\widetilde{v}_{22}(C\widetilde{v}_{22}+9\hpe\widetilde{v}_{32})
\Big]\label{gint}
\ee
with   
\beq
C \equiv 12 \get^\perp \gc - 6 \get^\parallel \get^\perp
	+ 6 (\get^\parallel)^2 \get^\perp + 6 (\get^\perp)^3
- 2 \get^\perp \gx^\parallel- 2 \get^\parallel \gx^\perp
	- \sqrt{\frac{\e}{2}}\frac{1}{\kappa H^2}(W_{211} + W_{222}),
\label{Ci}
\eeq
where $W_{mnl}=W^{,ABC}e_{mA}e_{nB}e_{lC}$. 
It is from this integrated effect that any large, persistent non-Gaussianity 
originates, if we consider only models where the isocurvature modes have
vanished by the end of inflation.
For the analytical approximations that we will provide (in addition to the
exact numerical results), it is useful to note that within the slow-roll
approximation $g_{int}$ can be rewritten as 
\be
g_{int}(k_1,k_2) = \bv_{12k_1} G_{22 k_1 k_2} \left( -\eta^\perp_{k_1} 
+ \frac{(\epsilon_{k_1}
+\eta^\parallel_{k_1} -\chi_{k_1})\chi_{k_1}}{2\eta^\perp_{k_1}} \right) 
+ \tg_{int}(k_1,k_2),
\label{tgint}
\ee
where $\tg_{int}$ is another integral that is identically zero for the two-field
quadratic model, or even more generally for any two-field equal-power sum model 
(see \cite{Tzavara:2010ge} for details).

\subsection{Discussion}

\begin{figure}
\begin{tabular}{cc}
\includegraphics[scale=0.8]{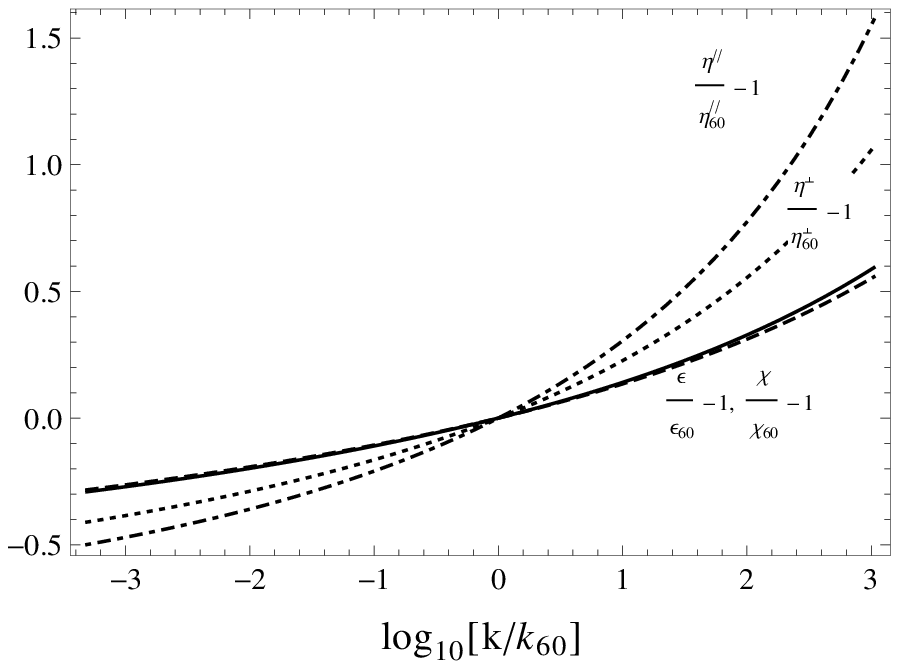}
&\includegraphics[scale=0.85]{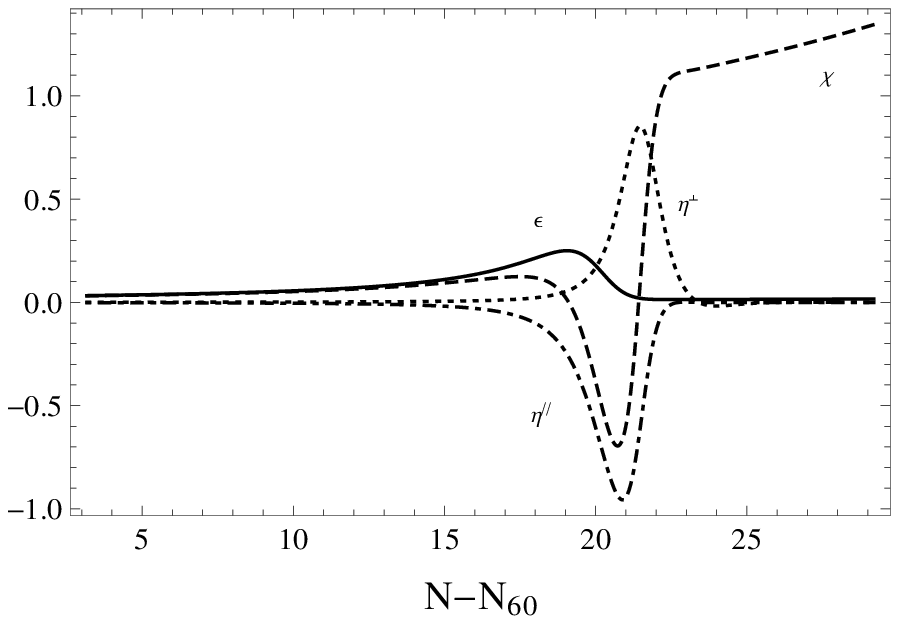}
\end{tabular}
\caption{Left: The relative change of the horizon-crossing first-order slow-roll 
parameters $\e$ (solid curve), $\hpa$ (dot-dashed curve), $\hpe$ (dotted curve) and $\chi$ (dashed curve) 
at $t_k$ as a function of the 
ratio $k/k_{60}$ of the horizon-exit scale to the scale that left the horizon $60$ e-folds 
before the end of inflation, 
for the model (\ref{qua}) with mass ratio $m_{\phi}/m_{\gs}=9$. 
Right: The evolution of the first-order slow-roll parameters $\e$ (solid curve), $\hpa$ (dot-dashed curve), $\hpe$ (dotted curve) and $\chi$ (dashed curve) as a function of the number of e-foldings $N-N_{60}$ for the time interval around the turning of the fields, for the same model.
}
\label{fig2}
\end{figure}

Inspecting (\ref{fNLgeni}) ones sees that there are two sources of momentum 
dependence for $\f$: the slow-roll parameters 
at horizon-crossing and the Green's functions $G_{ab}$ or their combinations 
$\bv_{a2}$. In order to study their impact 
we shall use the quadratic model 
\be
W=\frac{1}{2}m_{\phi}^2\phi^2+\frac{1}{2}m_{\gs}^2\gs^2,\label{qua}
\ee
with $m_{\phi}/m_{\gs}=9$. The procedure to follow is to solve for the 
background quantities (\ref{fieldeq}) and then for the Green's functions (see subsection 2.2 in \cite{Tzavara:2010ge}) 
in order to apply the formalism. The quadratic model's Green's functions can be found 
numerically, or even analytically within the slow-roll approximation, which is valid for a 
small mass ratio like the one we chose here. However, all our calculations in this paper 
are numerical and exact, without assuming the slow-roll approximation after horizon 
crossing. We only use the slow-roll approximation after horizon crossing for the
analytical approximations that we provide (e.g.\ eq.~(\ref{fnlfin})) and 
sometimes to clarify the physical interpretation of results (e.g.\ the use of 
(\ref{difu}) below to explain the behaviour of $\bv_{12}$).
Inflation ends at $t_f$ defined as the time when $\e_f=1$. 
From now on a subscript $f$ 
will denote quantities evaluated at the end of inflation. 
We also define the scale that exited the 
horizon $60$ e-folds before the end of inflation as $k_{60}$ and use it as a reference 
scale, around which we perform our computations ($k_{60}$ being the scale that 
corresponds to the text books' minimal necessary amount of inflation). 

In figure \ref{fig2} we plot the first-order slow-roll parameters for a 
range of horizon-crossing times around $k_{60}$. While the heavy field rolls 
down its potential the slow-roll parameters increase, reflecting the 
evolution of the background. This implies that 
$\f$, which is in general proportional to the 
slow-roll parameters evaluated at $t_k$ and $t_{k'}$, 
should increase as a function of $k$ and $k'$. 
This can easily be verified for the initial value of $f_{\mathrm{NL},in}$ at 
$t=t_{k'}$, which according to (\ref{fNLgeni}) with $\bv_{12k'}=0$ takes 
the value 
\be
-\frac{6}{5}f_{\mathrm{NL},in}=\e_{k'}+\hpa_{k'}+\frac{2\frac{\gamma_k^2}{\gamma^2_{k'}}G_{12k'k}}{1+2\frac{\gamma_k^2}{\gamma^2_{k'}}
[1+(G_{12k'k})^2]}\hpe_{k'}G_{22k'k}.
\label{fnlin}
\ee

\begin{figure}
\begin{tabular}{cc}
\includegraphics[scale=0.8]{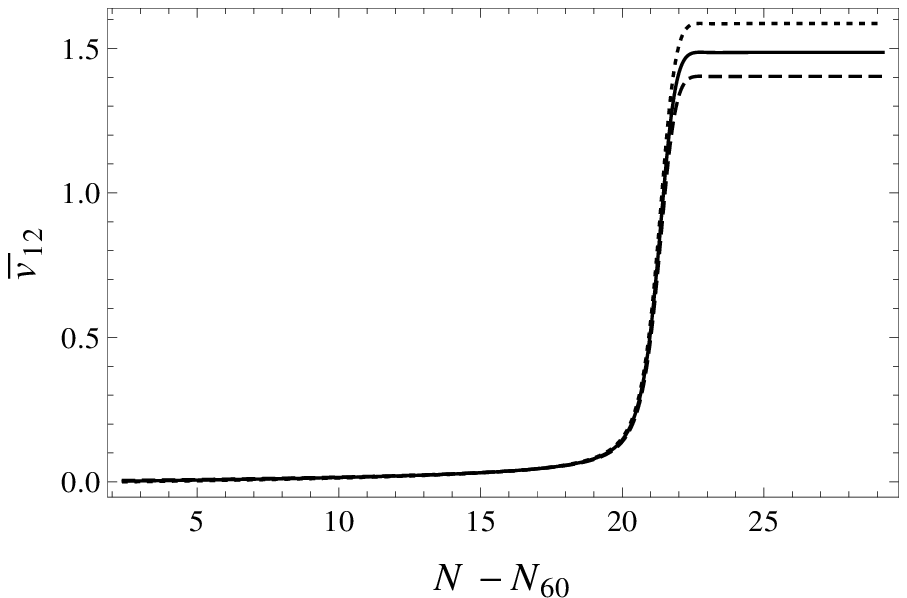}
&\includegraphics[scale=0.8]{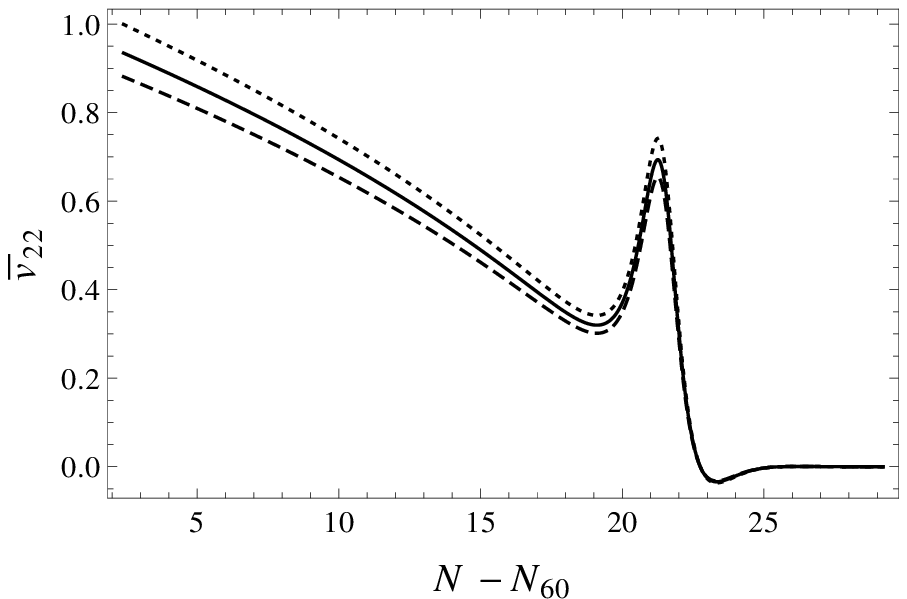}
\end{tabular}
\caption{The evolution of the transfer functions $\bv_{12}$ (left) and 
$\bv_{22}$ (right) as a function of the number 
of e-foldings $N-N_{60}$ for the time interval around the turning of the fields 
and for different horizon exit scales, varying 
from top to bottom as $k_{60}\times10$ (dotted curve), $k_{60}$ (solid curve) and $k_{60}/10$ (dashed curve), 
for the model (\ref{qua}) with mass ratio $m_{\phi}/m_{\gs}=9$.}
\label{fig1}
\end{figure}

Apart from the slow-roll parameters the other source of momentum dependence for $\f$ lies in the Green's functions and 
particularly how their time evolution depends on the relevant horizon-crossing scale. The two main quantities that we 
need to study in order to understand their impact on $\f$ are the transfer functions $\bv_{12}$ and $\bv_{22}$. This is due to the fact 
that $\bv_{32}$ is slow-roll suppressed and the rest of the Green's functions appearing in (\ref{fNLgeni}) can 
be rewritten in terms of $\bv_{12}$ and $\bv_{22}$ within the slow-roll approximation (for details, see \cite{Tzavara:2010ge}).  
In particular  $G_{a3}=G_{a2}/3$, $G_{32}(t,t_k)=-\chi(t) G_{22}(t,t_k)$ and hence $G_{a2} \approx \bv_{a2}$. Note that except for the era of 
the turning of the fields, the slow-roll assumption is a good approximation during inflation in this particular model.   
The slow-roll evolution equations for $\bv_{12k}$ and $\bv_{22k}$ are
\be
\frac{\d}{\d t}\bv_{12k}=2\hpe\bv_{22k}\qquad\mathrm{and}\qquad\frac{\d}{\d t}\bv_{22k}=-\chi\bv_{22k}.\label{difu}
\ee

As was discussed above, $\bv_{12}$ describes how the isocurvature mode sources the 
adiabatic one, while $\bv_{22}$ describes how the 
isocurvature mode sources itself.  
By definition $\bv_{12}(t_k,t_k)=0$ and $\bv_{22}(t_k,t_k)=1$ at horizon crossing, 
since no interaction of the different modes has yet occurred (see also 
(\ref{green}) and (\ref{Green_norm})). 
For the transfer functions of the adiabatic mode one finds that $\bv_{11}=1$ 
and $\bv_{21}=0$, since the curvature perturbation is conserved for purely 
adiabatic perturbations and adiabatic perturbations cannot source entropy 
perturbations. 
In order to better understand the 
role of the transfer functions, we can use the Fourier transformation of the 
perturbations \cite{Tzavara:2010ge} along with these last identities, to find
\bea
&&\zeta_{1}(t)=\int\frac{\d^3\vc{k}}{(2\pi)^{3/2}}\gamma_k\bv_{1m}\hat{a}^{\dagger}_m(\vc{k})
 e^{\rmi \vc{k}\cdot \vc{x}}=\zeta_{1}(t_k)+\bv_{12}(t,t_k)\zeta_{2}(t_k),\nn\\
&&\zeta_{2}(t)=\int\frac{\d^3\vc{k}}{(2\pi)^{3/2}}\gamma_k\bv_{2m}\hat{a}^{\dagger}_m(\vc{k})
 e^{\rmi \vc{k}\cdot \vc{x}}=\bv_{22}(t,t_k)\zeta_{2}(t_k),\label{ze}
\eea
where $\zeta_{m}$ with $m=1,2$ are the first-order adiabatic and isocurvature 
perturbation. 
 
Let us start by discussing the time evolution of $\bv_{12}$. Each one of the 
curves on the left-hand side of figure \ref{fig1} corresponds to the 
time evolution of $\bv_{12}$ for a different horizon-exit scale. 
At $t=t_k$, i.e.\  when the relevant mode $k$ exits the horizon, $\bv_{12k}=0$   
since the isocurvature mode has not had time to affect the adiabatic one. Outside 
the horizon and well in the slow-roll regime of the sole dominance of the heavy field, the
isocurvature mode sources the adiabatic one and the latter slowly increases. 
As time goes by, the heavy field rolls down its potential 
and the light field becomes more important. 
During this turning of the field trajectory, the slow-roll parameters suddenly
change rapidly, with important consequences for the evolution of the adiabatic
and isocurvature mode. The transfer function
$\bv_{12k}$ grows substantially during that era because of the increasing 
values of $\hpe$ in (\ref{difu}) as well as the growing contribution of $\bv_{22k}$, 
to become constant afterwards  
when the light field becomes dominant in an effectively single-field universe. 

Note that the earlier the mode exits the horizon, the smaller is the final $\bv_{12k}$. This is opposite to the behaviour of the initial 
value, just after horizon-crossing, 
when the earlier the scale exits the horizon the more has its adiabatic mode been sourced by the isocurvature one at a given time $t$, and hence the 
larger is its $\bv_{12k}$. 
This can be understood by the evolution equations of $\bv_{12k}$ and $\bv_{22k}$ 
in (\ref{difu}), 
showing that $\bv_{12k}$ is sourced by $\bv_{22k}$, which itself is a decreasing function 
of time, 
at least during eras when the universe is dominated by a single field (see 
the right-hand side of figure \ref{fig1}). If the equation (\ref{difu}) for $\bv_{12k}$ did not depend on $\hpe$, the $\bv_{12k}$ curves 
would never cross each other since they would be similar and only boosted by their horizon-crossing time shift. It is the increasing 
value of $\hpe$ 
that results in the larger values of $\bv_{12k}$ for larger $k$.

\begin{figure}
\begin{tabular}{cc}
\includegraphics[scale=0.8]{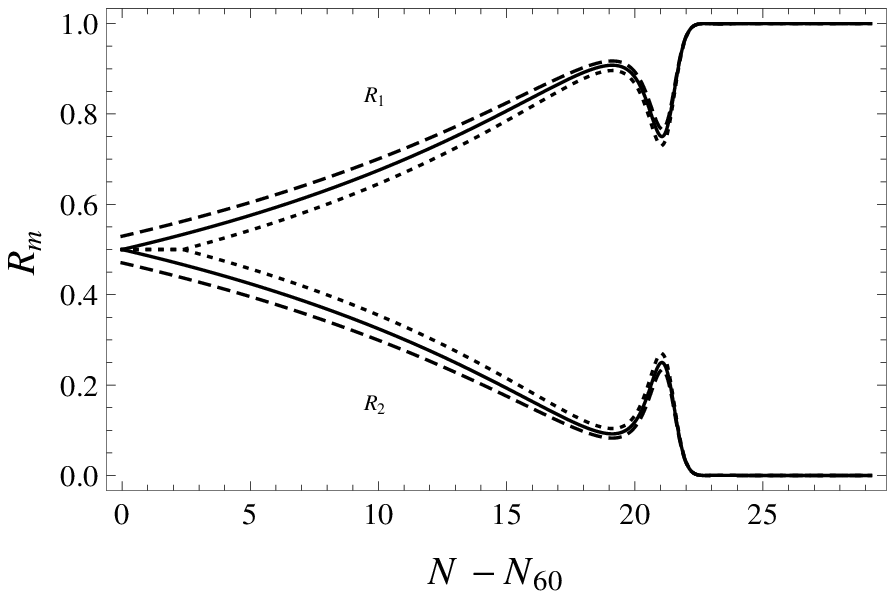}
&\includegraphics[scale=0.85]{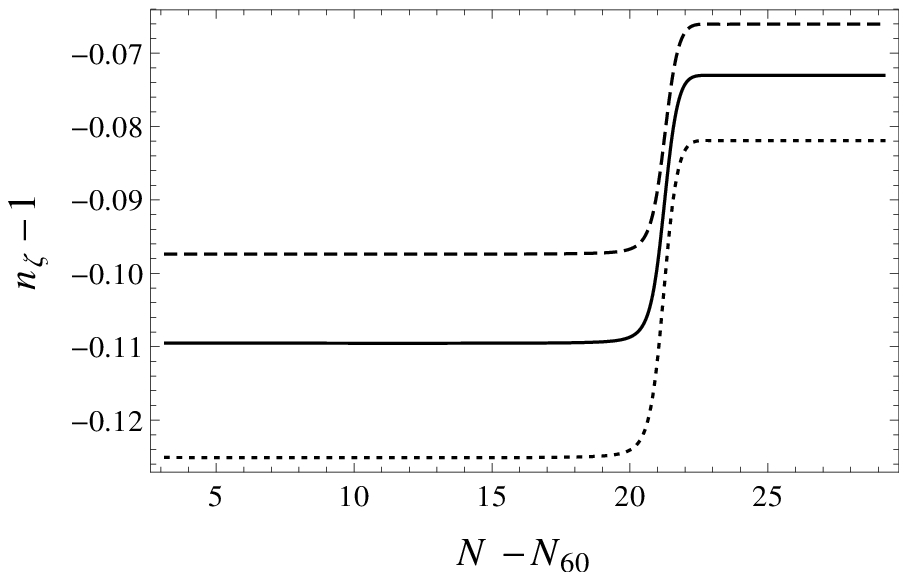}
\end{tabular}
\caption{Left: The time evolution of the adiabatic $R_1$ and the isocurvature 
$R_2$ ratios of power spectra as a function of the number of e-foldings $N-N_{60}$ 
for scales $k_{60}/10$ (dashed line), $k_{60}$ (solid line) and 
$k_{60}\times10$ (dotted line). Right: The time evolution of the spectral index (\ref{spectral}) 
as a function of the number of e-foldings $N-N_{60}$ 
for scales $k_{60}/10$ (dashed line), $k_{60}$ (solid line) and 
$k_{60}\times10$ (dotted line). Both plots are made 
for the model (\ref{qua}) with mass ratio $m_{\phi}/m_{\gs}=9$.}
\label{fig11}
\end{figure}

On the right-hand side of figure \ref{fig1} we show the evolution of $\bv_{22}$. 
According to (\ref{difu}) $\bv_{22}$ and hence the isocurvature mode evolves independently from the adiabatic mode. 
At horizon-crossing $t=t_k$, the transfer function $\bv_{22k}=1$.
Once outside the horizon, the isocurvature mode decays due to the small but positive values of $\chi$.
During the turning of the fields the slow-roll parameters evolve rapidly, thus leading to first an enhancement of $\bv_{22k}$ and then a diminution due to the varying value of $\chi$ in (\ref{difu}). 
As can be seen from the right-hand side plot in figure \ref{fig2}, during the turning $\chi$ first becomes negative and then positive. 
After the turning of the fields, the remnant isocurvature modes again decay and (for this model) at the end of inflation none are left.
The parameter $\chi$ plays a crucial role in the evolution of the isocurvature mode. 
It represents effectively the second derivative of the potential in the $22$ direction. 
Before the turning of the fields the trajectory goes down the potential in the 
relatively steep $\phi$ direction, which means that $W_{22}$ then corresponds
with the relatively shallow curvature in the direction of the light field 
$\sigma$ and hence $\chi$ is small. After the turning the trajectory goes along
the bottom of the valley in the $\sigma$ direction and $W_{22}$ corresponds with
the large curvature of the potential in the perpendicular direction, leading 
to large values of $\chi$. The negative values of $\chi$ during the turn come
from the contribution of $\hpa$.

Instead of looking at the tranfer functions, using (\ref{ze}) one can also 
construct more physical quantities from the operators $\zeta_{m}$ 
and hence from the $\bv_{m2}$, namely the ratios 
of the adiabatic and isocurvature power spectrum to the total power spectrum:
\bea
&&R_1\equiv\frac{\langle\zeta_{1}\zeta_{1}\rangle}{\langle\zeta_{1}\zeta_{1}\rangle+
\langle\zeta_{2}\zeta_{2}\rangle}=\frac{1+\lh\bv_{12}\rh^2}{1+\lh\bv_{12}\rh^2+\lh\bv_{22}\rh^2},\nn\\
&&R_2\equiv\frac{\langle\zeta_{2}\zeta_{2}\rangle}{\langle\zeta_{1}\zeta_{1}\rangle+
\langle\zeta_{2}\zeta_{2}\rangle}=\frac{\lh\bv_{22}\rh^2}{1+\lh\bv_{12}\rh^2+\lh\bv_{22}\rh^2}.
\eea
These are plotted on the left-hand side of figure \ref{fig11} as a function 
of the number of e-foldings for different scales. 
One can clearly see that both 
ratios start as equal to $1/2$ when the scale exits the horizon, while 
afterwards the adiabatic ratio $R_1$ increases to reach $1$ at the end 
of inflation and the isocurvature $R_2$ decreases to reach $0$, for this 
particular model. During the turning of the fields we see that the temporary
increase in the isocurvature mode due to the negative value of $\chi$ is 
reflected in $R_2$, while the adiabatic $R_1$ necessarily has the opposite 
behaviour. 

On the right-hand side of figure \ref{fig11} we plot the time evolution of the spectral index (\ref{spectral}) of the power spectrum. 
The spectral index measures by construction the tilt of the power spectrum for different horizon-crossing scales and hence it 
depends on the horizon-crossing slow-roll parameters. For multiple-field models the power spectrum evolves during inflation even after horizon-crossing, 
and so does the spectral index.   
During the turning of the fields the spectral index increases, to remain constant afterwards. 
The earlier 
a scale exits the horizon the less negative is its spectral index $n_{\zeta}-1$. This implies that the power spectrum itself decreases faster 
for larger horizon-crossing scales. This is due to the fact that except for the factor $1+(\bv_{12k})^2$ in the expression for the power 
spectrum there is also an inverse power of $\e_k$ (see \cite{Tzavara:2010ge}).

\subsection{Spectral indices}

Finally let us discuss the scale dependence of the local $\f$ in terms of the relevant spectral indices. 
Equation (\ref{fNLgeni}) for an isosceles triangle implies that 
\bea
-\frac{6}{5}\f&=&\frac{1}{(2\pi^2)^2}\frac{f(k',k')+2(\frac{k'}{k})^3f(k',k)}
{\mathcal{P}_\zeta(k')^2+2(\frac{k'}{k})^3\mathcal{P}_\zeta(k')\mathcal{P}_\zeta(k)},
\eea
where 
\be
f(k',k) = -2 (k'k)^3 \gamma_{k'}^2 \gamma_k^2 \, \bv_{12k'} \bv_{12k}
\left( g_{sr}(k',k) + g_{iso}(k',k) + g_{int}(k',k) \right).
\ee 
For an arbitrary triangle configuration this is generalized as
\bea
-\frac{6}{5}\f&=&\frac{1}{(2\pi^2)^2}\frac{k_3^3f(k_1,k_2)+\mathrm{perms.}}
{k_3^3\mathcal{P}_\zeta(k_1)\mathcal{P}_\zeta(k_2)+\mathrm{perms.}}.
\eea
The local $\f$ depends on a two-variable function $f(k_1,k_2)$, with $k_1\geq k_2$. This is due to its super-horizon origin, which 
yields classical non-Gaussianity proportional to products of two power spectra. 
Hence one expects that the scale dependence of 
$\f$ can be expressed in terms of only two spectral indices, characterizing the function $f$. 
Notice that this is particular to the local case. In general the bispectrum 
cannot be split as a sum of two-variable functions and one anticipates that three spectral indices would be needed. 

The next issue to be resolved is which are the relevant spectral indices for $f$. The naive guess would be 
$f(k_1,k_2)=f(k_{1,0},k_{2,0})(k_1/k_{1,0})^{\tn_{k_1}}(k_2/k_{2,0})^{\tn_{k_2}}$. 
We tested this parametrization and we did not find good agreement with the exact value of $f$. 
Instead of that, we found that $f$ is best approximated by keeping either the shape or the magnitude of the triangle constant. 
This statement can be expressed as 
\be
f(k_1,k_2)=f_0\lh\frac{K}{K_0}\rh^{\tn_{K}}\lh\frac{\omega}{\omega_0}\rh^{\tn_{\omega}},
\label{ind}
\ee
where
\be
\tn_{K}\equiv\frac{\d\ln f}{\d\ln K}\qquad\mathrm{and}\qquad \tn_{\omega}\equiv\frac{\d\ln f}{\d\ln\omega}\label{inddef}
\ee
and
\be
\omega\equiv\frac{k_1}{k_2}=\frac{1+\beta}{2(1-\beta)}. 
\ee
The last equality is valid only for the isosceles case $\gamma=0$ (see
(\ref{newvar})). We dropped 
the $-1$ of the power spectrum spectral index definition to follow the 
definitions in \cite{Byrnes:2009pe,Byrnes:2010ft}. 
We added a tilde to indicate that these spectral indices are defined for the
function $f$, not yet for the full $\f$.
In the next two sections we are going to examine the scale-dependence of $\f$, 
changing the magnitude and the shape of the triangle separately, 
and verify assumption (\ref{ind}).

\section{Changing the magnitude of the triangle}\label{conf}

\begin{figure}
\begin{center}
\includegraphics[scale=0.8]{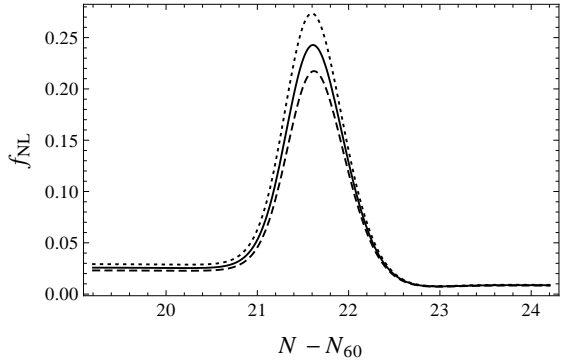}
\end{center}
\caption{The time evolution of $\f$ in terms of the number of e-foldings 
$N-N_{60}$ around the time of the turning of the fields,
for equilateral ($\omega=1$) triangles with $K=(3/2)k_{60}$ (solid curve), 
$K=(3/2)k_{60}/10$ (dashed curve) and $K=(3/2)k_{60}\times10$ 
(dotted curve),
for the model (\ref{qua}) with mass ratio $m_{\phi}/m_{\gs}=9$.}
\label{fig56}
\end{figure}

\begin{figure}
\begin{tabular}{cc}
\includegraphics[scale=0.8]{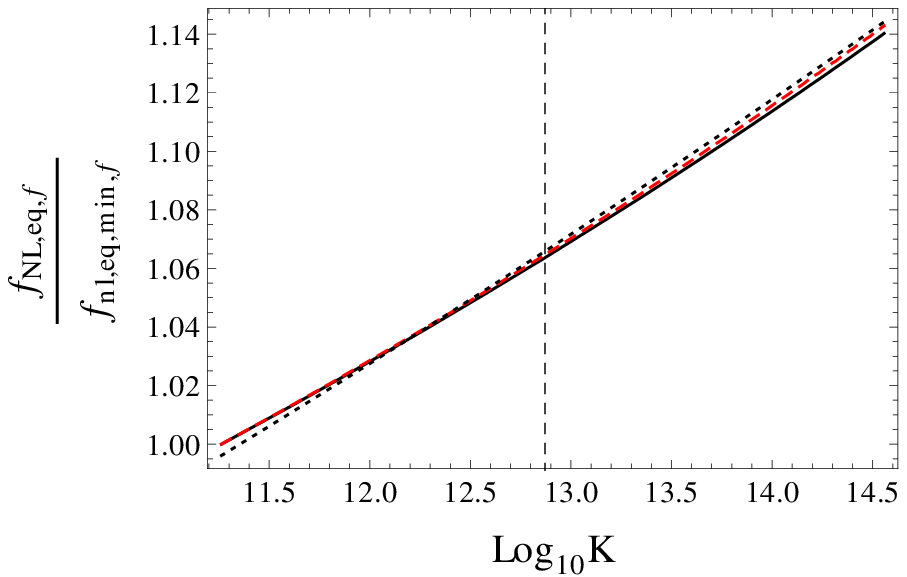}
&\includegraphics[scale=0.8]{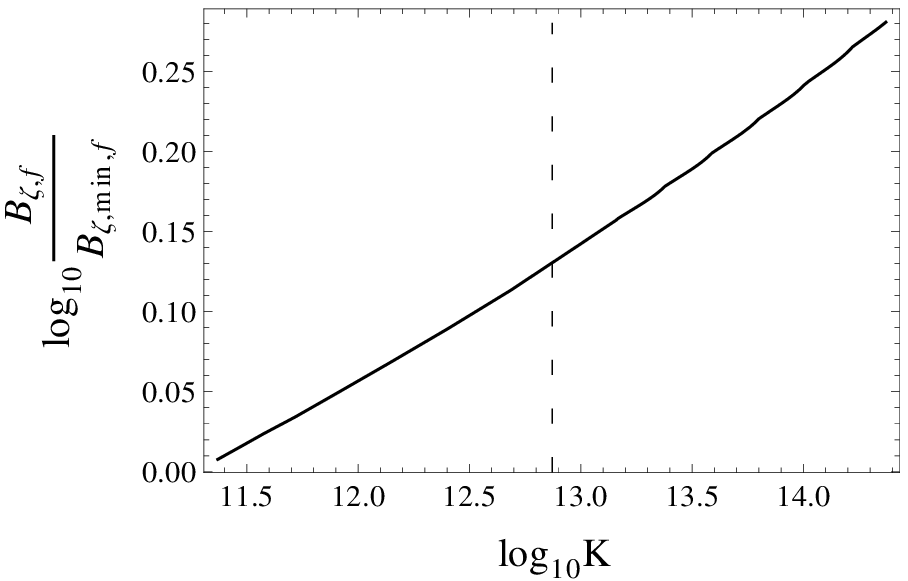}
\end{tabular}
\caption{Left: The relative change of the final value of $\f$ arbitrarily normalized to one at the smallest value of $K$ on the figure, 
as a function of $K$ for equilateral triangles ($\omega=1$), calculated exactly (solid curve), using the analytical approximation (\ref{fnlfin}) (dashed red curve) 
and using the shape index (\ref{nkeq}) (dotted black curve).  
Right: The logarithm of the final value of the exact bispectrum, similarly normalized,  
as a function of 
$K$ for equilateral triangles ($\omega=1$).  
Both figures are for the quadratic model (\ref{qua}) with mass ratio 
$m_{\phi}/m_{\gs}=9$. The vertical dashed line corresponds to $K=(3/2)k_{60}$.}
\label{fig31}
\end{figure}

In this section we shall study the behaviour of $\f$ for triangles 
of the same shape but different size, see the left-hand side of 
figure~\ref{fig01}. In figure 
\ref{fig56} we plot the time evolution of $\f$ for equilateral 
triangles (the result would remain qualitatively the same for any isosceles 
triangle) of perimeter $K=(3/2)k_{60}\times10$ (top curve), 
$K=(3/2)k_{60}$ (middle curve) and $K=(3/2)k_{60}/10$ (bottom curve). The later 
the relevant scale exits the horizon the larger is its initial $\f$ as 
explained in the previous section. $\f$ grows during the turning of the 
fields due to isocurvature effects as described by (\ref{giso}) and 
(\ref{gint}), but by the end of inflation, when isocurvature modes vanish, it 
relaxes to a small, slow-roll suppressed value (see e.g.\  \cite{Vernizzi:2006ve,Tzavara:2010ge}). 
In figure \ref{fig31} we plot the final value of $\f$ (left) and the final 
value of the bispectrum (right) for equilateral triangles, varying $K$ for 
values around $K=(3/2)k_{60}$, within the Planck satellite's resolution 
($k'/k\sim1000$). The later the scale exits the horizon, i.e.\  the larger
$K$, the larger is the final value of $\f$ and of the bispectrum. 

The final value of $\f$ can be found analytically for the quadratic model 
within the slow-roll approximation. 
By the end of inflation $\bv_{22,f}=0$ so that 
(\ref{giso}) vanishes, while (\ref{gint}) can be further 
simplified to give some extra horizon-crossing terms and a new integral 
$\tg_{int}$ that is identically zero for the quadratic potential 
(see (\ref{tgint}) and \cite{Tzavara:2010ge}). 
For simplicity we give here the final value of 
$\f$ for equilateral triangles, 
\be
f_{\mathrm{NL},eq,f}(k)=\frac{
3\lh\bv_{12k}\rh^2\lh\e_k+\hpa_k-\chi_k+\frac{\hpe_k}{\bv_{12k}}\rh
+\lh\bv_{12k}\rh^3\lh\hpe_k-
\frac{\lh\e_k+\hpa_k-\chi_k\rh\chi_k}{\hpe_k}\rh+\e_k+\hpa_k}
{\lh1+\lh\bv_{12k}\rh^2\rh^2}.\label{fnlfin}
\ee
This formula is actually valid for any two-field model for which isocurvature modes
vanish at the end of inflation and for which $\tg_{int}=0$, 
like for example equal-power sum models. 
Inspecting the various terms it turns out that although $\bv_{12k}$ tends to decrease 
the value of $f_{\mathrm{NL},f}$ as a function of $k$, it is the contribution of the horizon-crossing  
slow-roll parameters that wins and leads to an increase of the parameter 
of non-Gaussianity for larger horizon-crossing scales.
Note that for equilateral triangles $K$ is simply $3k/2$.

\begin{figure}
\begin{tabular}{ll}
\includegraphics[scale=0.8]{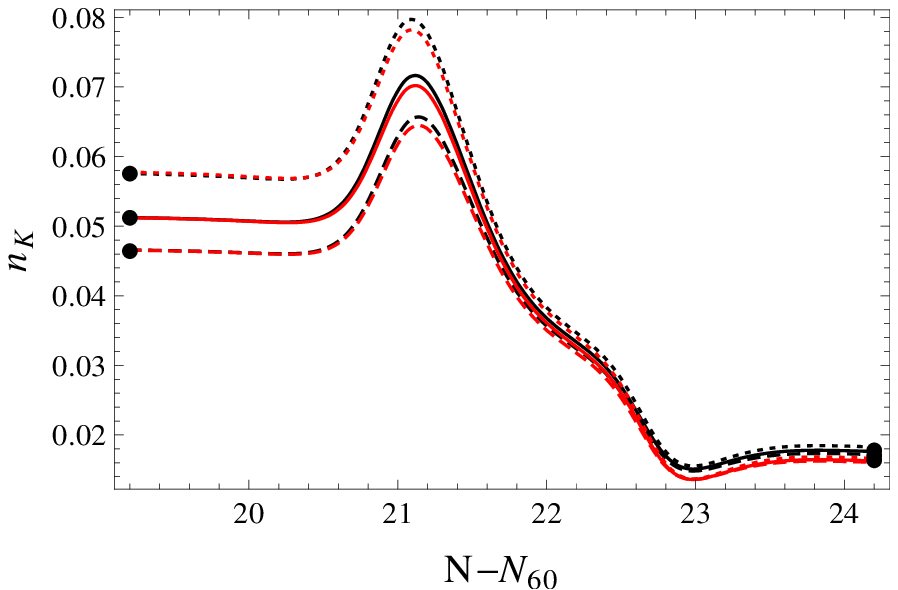}
&\includegraphics[scale=0.8]{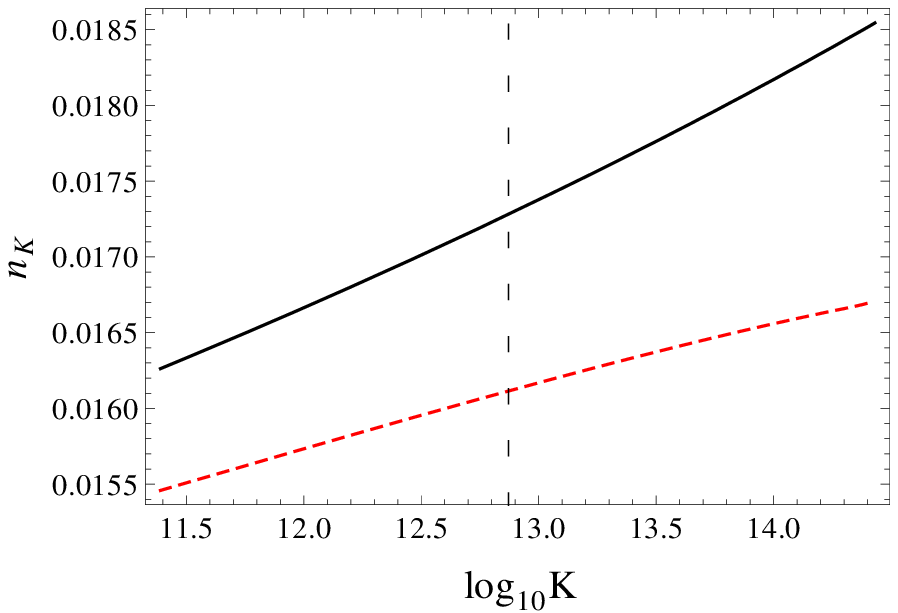}
\end{tabular}
\caption{Left: The time evolution of the conformal index 
$n_K$ (\ref{con}) around the time of the turning of the fields,
for triangles with $\omega=1$ (black curves) and 
$\omega=5/2$ (red curves, below the $\omega=1$ curves), 
with perimeter $K=(3/2)k_{60}$ (solid curve), 
$K=(3/2)k_{60}/10$ (dashed curve) and $K=(3/2)k_{60}\times10$ (dotted curve). 
The three points on the left and on the right correspond to the analytical values 
of the index as calculated from (\ref{ind1}) and (\ref{A1}) respectively. 
Right: The final value of the conformal index $n_K$ 
(\ref{con}) for triangles with $\omega=1$ (black curve) and $\omega=5/2$ (red dashed curve)  
as a function of $K$. 
Both figures are for the model (\ref{qua}) with mass 
ratio $m_{\phi}/m_{\gs}=9$.}\label{fig51}
\end{figure}

We turn now to the spectral index $n_{K}$. Using (\ref{inddef}) with (\ref{par}) and assuming 
that $\gamma=0$ and $\beta=\mathrm{const.}$, we can express $\tn_{K}$ in terms of the horizon-crossing time derivatives as 
\be 
\tn_{K}(t;t_{k_1},t_{k_2})=\frac{\partial\ln{f}}{\partial t_{k_1}}\frac{1}{1-\e_{k_1}}
+\frac{\partial\ln{f}}{\partial t_{k_2}}\frac{1}{1-\e_{k_2}}.
\label{con}
\ee 
Then $\f$ takes the form
\be
-\frac{6}{5}\f=\frac{1}{(2\pi^2)^2}\frac{f(k'_0,k'_0)\lh\frac{K}{K_0}\rh^{\tn_{K}(t_{k'_0},t_{k'_0})}+
2\omega^3f(k'_0,k_0)\lh\frac{K}{K_0}\rh^{\tn_{K}(t_{k'_0},t_{k_0})}}
{\mathcal{P}_\zeta(k'_0)^2\lh\frac{k'}{k'_0}\rh^{2(n_\zeta(t_{k'_0})-1)}
+2\omega^3\mathcal{P}_\zeta(k'_0)\mathcal{P}_\zeta(k_0)\lh\frac{k'}{k'_0}\rh^{n_\zeta(t_{k'_0})-1}\lh\frac{k}{k_0}\rh^{n_\zeta(t_{k_0})-1}}.
\label{nkf}
\ee
Note that the ratios $k'/k'_0=k/k_0=K/K_0$, since $\beta=\mathrm{const}$. 

The above formula can be simplified in the limit of squeezed-triangle configurations, as well as in the equilateral limit. 
When one takes the squeezed limit $\omega^3\mg1$ (note that this would be true for $\beta\gtrsim2/3$), one finds:
\bea
-\frac{6}{5}\f&=&\frac{1}{(2\pi^2)^2}\frac{f(k'_0,k_0)}{\mathcal{P}_\zeta(k'_0)\mathcal{P}_\zeta(k_0)}
\lh\frac{K}{K_0}\rh^{\tn_{K}(t_{k'_0},t_{k_0})-n_\zeta(t_{k'_0})-n_\zeta(t_{k_0})+2}\nn\\
&\equiv&-\frac{6}{5}f_{\mathrm{NL},0}
\lh\frac{K}{K_0}\rh^{n_{K}(t_{k'_0},t_{k_0})},
\eea
where
\be
n_K(t;t_{k'},t_k)\equiv\frac{\d \ln\f}{\d\ln K}=\frac{\partial\ln{\f}}{\partial t_{k'}}\frac{1}{1-\e_{k'}}
+\frac{\partial\ln{\f}}{\partial t_{k}}\frac{1}{1-\e_{k}}.\label{nkt}
\ee
For the equilateral case $\omega=1$, (\ref{nkf}) becomes
\bea
-\frac{6}{5}\f&=&\frac{1}{(2\pi^2)^2}\frac{f(k'_0,k'_0)}{\mathcal{P}_\zeta(k'_0)^2}
\lh\frac{K}{K_0}\rh^{\tn_{K}(t_{k'_0},t_{k'_0})-2n_\zeta(t_{k'_0})+2}\nn\\
&\equiv&-\frac{6}{5}f_{\mathrm{NL},0}
\lh\frac{K}{K_0}\rh^{n_{K}(t_{k'_0},t_{k'_0})}.\label{nkeq}
\eea

The conformal spectral index $n_K$ measures the change of $\f$ due to the 
overall size of the triangle, namely due to a conformal transformation of the triangle. 
For an isosceles triangle this is conceptually sketched on the left-hand side of 
figure \ref{fig01}, but it can be generalized for any shape. $n_K$ coincides with the 
$n_{\f}$ of \cite{Byrnes:2009pe,Byrnes:2010ft} and grossly speaking it describes the 
tilt of $\f$ due to the pure evolution of the inflationary background (note that for 
an equilateral triangle this statement would be exact).

On the left-hand side of figure \ref{fig51} we plot the time evolution of the conformal spectral index for an equilateral 
$\omega=1$ and an isosceles $\omega=5/2$ triangle that exited the horizon 
at three different times, namely for $K=(3/2)k_{60}$ (solid curve), 
$K=(3/2)k_{60}/10$ (dashed curve) and $K=(3/2)k_{60}\times10$ (dotted curve). We plot 
the $\omega=5/2$ case only to demonstrate that the results remain qualitatively the same; 
we shall study the effect of different triangle shapes in the next section. The characteristic peaks that $n_K$ exhibits during the 
turning of the fields are inherited from the behaviour of $\f$ at that time and it is a new feature that is absent 
in the time evolution of the power spectrum spectral 
index $n_{\zeta}-1$ (see the right-hand side of figure \ref{fig11}).

In the context of the long-wavelength formalism we are 
restricted to work with the slow-roll approximation at horizon 
exit, so that the slow-roll parameters at that time should be small and vary 
just a little. This should be reflected in the initial value of 
the spectral index, which should be $\mathcal{O}(\e_{k'})$. 
In another paper we will study models that do not necessarily satisfy this constraint, 
using the exact cubic action derived in \cite{Tzavara:2011hn} instead of the 
long-wavelength formalism. 
The earlier 
the scale exits, e.g.\  the dashed curve, the smaller are the slow-roll 
parameters evaluated at horizon crossing and hence the 
smaller is the initial $n_K$.  
Indeed, using the definition (\ref{nkt}) with (\ref{fnlin}) for the initial 
value of $f_{\mathrm{NL},in}$, we find for equilateral triangles
\be
n_{K,in}=\frac{2\e_k^2+3\e_k\hpa_k+(\hpe_k)^2-(\hpa_k)^2+\xi^{\parallel}_k}
{\e_k+\hpa_k},\label{ind1}
\ee
which confirms the above statement. 

We notice that the initial, horizon-crossing, differences between the values of $n_K$ for the different horizon-crossing scales
mostly disappear by the end of inflation, after peaking during the turning of the fields. 
The final value of the spectral index is plotted on the right-hand side of figure 
\ref{fig51} and is smaller than its initial value. 
It exhibits a small running of $\mathcal{O}(10\%)$ within the 
range of scales studied, inherited from the initial dispersion of its values at horizon-crossing. 
To verify that $n_K$ describes well the behaviour of $\f$, we have plotted the
approximation (\ref{nkeq}) in figure \ref{fig31} where it can be compared with
the exact result.
We have also verified this for other inflationary models, including the potential 
\be
W=b_0-b_2\gs^2+b_4\gs^4+a_2\phi^2,\label{pot}
\ee
studied in 
\cite{Tzavara:2010ge}, able to produce $\f$ of ${\mathcal O}(1)$. 
The final value of the spectral index in that model is two orders of magnitude smaller than the value for the quadratic model. 
This is related essentially to the fact that for the potential (\ref{pot}) the turning of the fields, 
and hence the slow-roll breaking, occurs near the end of inflation. This means that at the horizon-crossing times of the scales 
of the triangle, slow-roll parameters change very slowly and as a result the initial variation of $\f$ is much smaller than the one 
for the quadratic potential. 
As a consequence, the final tilt of $\f$ will be smaller.

In Appendix \ref{app} one can find 
the analytical expression for the final value of the spectral index $n_K$, 
found by differentiating (\ref{fnlfin}). Since the 
result is rather lengthy and does not give any further physical intuition, we moved it to the appendix.

\section{Changing the shape of the triangle}\label{shape}

\begin{figure}
\begin{center}
\includegraphics[scale=0.8]{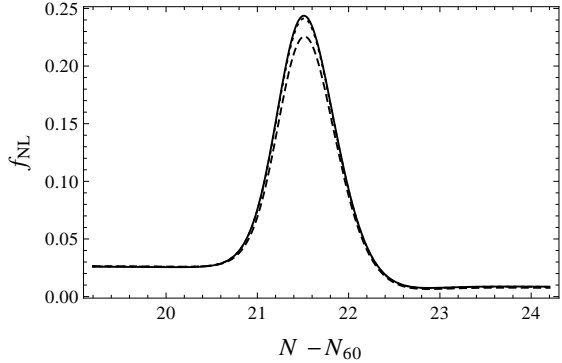}
\end{center}
\caption{The time evolution of $\f$ as a function of the number of e-foldings 
$N-N_{60}$ around the time of the turning of the fields,
for a triangle with $\omega=1$ (solid curve), $\omega=5/2$ (dotted curve) and $\omega=1000$ 
(dashed curve), all with fixed perimeter $K=3/2k_{60}$,
for the model (\ref{qua}) with mass ratio $m_{\phi}/m_{\gs}=9$.}
\label{fig55}
\end{figure}

After studying triangles with the same shape but varying size in the previous
section, we now turn to the scale dependence of $\f$ for triangles of the same perimeter but 
different shape, see the right-hand side of figure~\ref{fig01}. In figure \ref{fig55} we plot the time evolution of $\f$ during 
inflation for an equilateral $\omega=1$ (solid curve), an isosceles  
$\omega=5/2$ (dotted curve) and a squeezed $\omega=1000$ (dashed curve) triangle, 
all of perimeter $K=(3/2)k_{60}$, as a function of the number of e-foldings. The 
profile of the time evolution of $\f$ was discussed in the previous section. Here we 
are interested in the shape dependence of $\f$. 

\begin{figure}
\begin{tabular}{cc}
\includegraphics[scale=0.8]{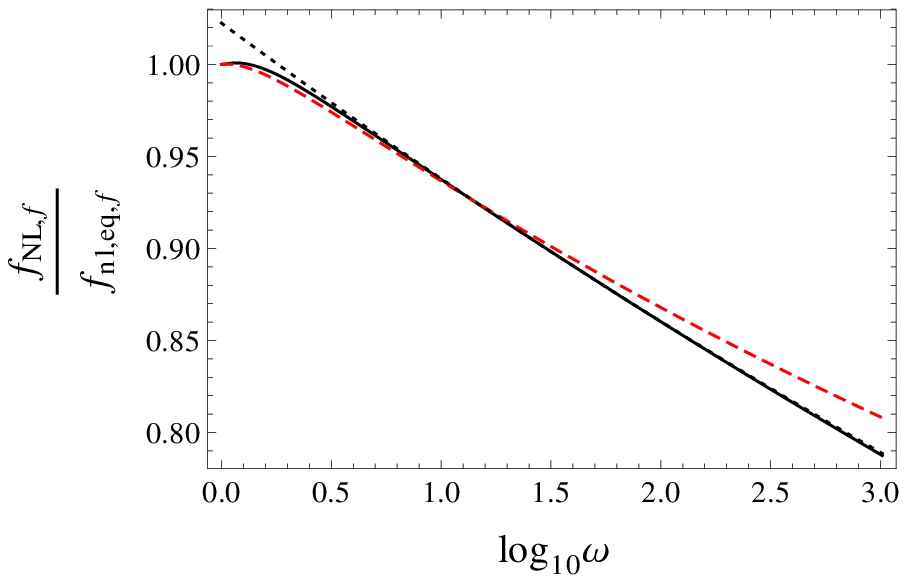}
&\includegraphics[scale=0.8]{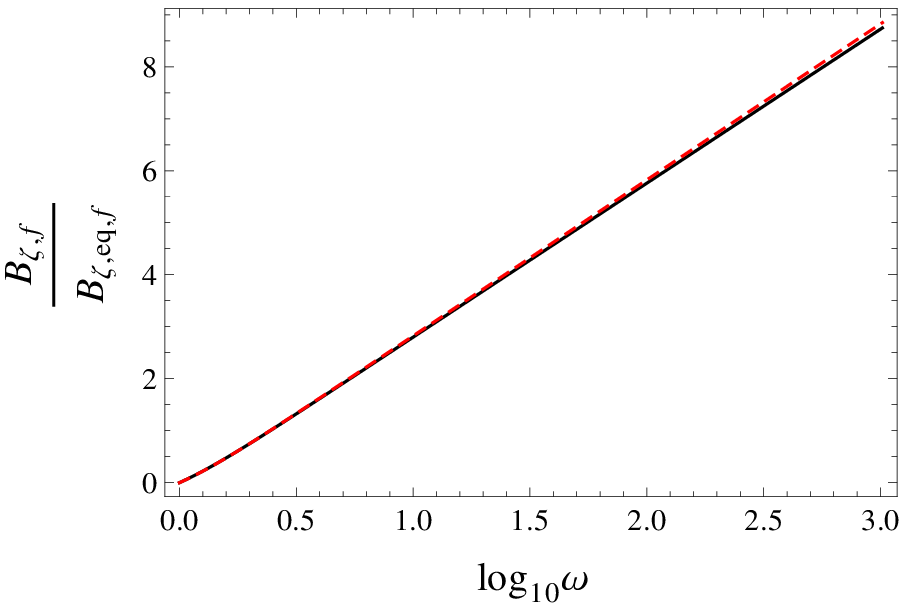}
\end{tabular}
\caption{Left: The final value of $\f$ normalised by the equilateral 
$\f$ as a function of $\omega$ for triangles with $K=
(3/2)k_{60}$ calculated exactly (black curve), using the analytical approximation (\ref{fnlb}) (dashed red curve) 
and using the shape index (\ref{into}) (dotted black curve). 
Right: The logarithm of the final value of the exact local bispectrum 
normalised by the local bispectrum computed on an equilateral triangle as a function of 
$\omega$ for triangles with $K=(3/2)k_{60}$ (black curve) and the same quantity assuming $\f$ scale independent (red dashed curve).  
Both figures are for the quadratic model (\ref{qua}) with mass ratio 
$m_{\phi}/m_{\gs}=9$.}
\label{fig3}
\end{figure}

Although it is during the peak that the variation of $\f$ for different shapes 
is more prominent, its final value is also affected. On the left-hand side of figure 
\ref{fig3} we plot the 
value of $\f$ at the end of inflation for triangles of perimeter $K=(3/2)k_{60}$, 
normalised by its value for the equilateral case ($\omega=1$), as a function
of $\omega$.  
The deviation of the values is small since it is related to 
horizon-exit slow-roll suppressed quantities. 
Within the long-wavelength formalism (or the $\delta N$ formalism) slow roll
at horizon crossing is a requirement. 
Nevertheless, the important conclusion here is that  
$\f$ decreases when the triangle becomes more squeezed. 
This can be attributed 
to the fact that the more squeezed is the triangle, the more the fluctuation $\zeta_k$ is 
frozen and behaves as part of the background when scale $k'$ crosses the horizon.  
As a result 
the correlation between $k$ and $k'$ becomes less and the resulting non-Gaussianity 
is smaller (see also the discussion below equation (\ref{fnlb}).)

An analytical formula can be found when applying the slow-roll approximation 
to expression 
(\ref{fNLgeni}) at the end of inflation, when isocurvature modes have vanished.  
We perform an integration by parts in the integral (see (\ref{tgint}) and 
\cite{Tzavara:2010ge}; as before $\tg_{int}=0$). 
More precisely, assuming that we are really 
in the squeezed limit $k\ll k'$, the ratio $\gamma_k^2/\gamma_{k'}^2$ becomes
very large and we can ignore the equilateral terms that depend only on $k'$
and not also on $k$. We also assume 
that the decaying mode has vanished to simplify the expressions for the Green's functions 
(see \cite{Tzavara:2010ge} and the discussion in section \ref{sour}). $G_{12k'k}$ can be set to zero as one can see in figure 
\ref{fig1} (since it is basically equal to $\bv_{12}$ and only involves times
at the very left-hand side of the figure). 
Moreover, the same figure shows that $\bv_{12k}/\bv_{12k'}\approx 1$ (in the
formula these ratios are always multiplied by slow-roll parameters, so that
the deviation from 1 would be like a second-order effect), so that we find 
in the end
\bea
f_{\mathrm{NL},sq,f}
&=&G_{22k'k}f_{\mathrm{NL},eq,f}(k')+\frac{1-G_{22k'k}}{\lh1+\lh\bv_{12k'}\rh^2\rh^2}
\Bigg[\e_{k'}+\hpa_{k'}+\lh\frac{2\hpe_{k'}}{\bv_{12k'}}-\chi_{k'}\rh\lh\bv_{12k'}\rh^2
\Bigg],\label{fnlb}
\eea
where $f_{\mathrm{NL},eq,f}$ is given in equation (\ref{fnlfin}). The only quantity in the above expression that depends on 
the shape of the triangle is $G_{22k'k}$, so it must be $G_{22k'k}$ that is responsible for the decreasing behaviour of $f_{\mathrm{NL},sq,f}$. 
Indeed, increasing $\omega$ for a constant perimeter $K$ 
of the triangle means increasing the interval $t_k-t_{k'}$ and hence decreasing the 
value of $G_{22k'k}$ (see the right-hand side of figure \ref{fig1}, since in the slow-roll regime $\bv_{22}=G_{22}$). 
This means that the interaction of the two modes becomes less important.   
In the complete absence of isocurvature modes $G_{22k'k}=0$ and $f_{\mathrm{NL},sq,f}$ takes its minimal value.  
It is only the isocurvature mode that interacts with itself and the 
greater is the difference between the two momenta the less is the interaction.  Notice that 
the single-field limit of this result would correspond to $G_{22k'k}=0$ and $\bv_{12}=0$. 
\footnote{Inspecting equation (\ref{fnlb}) we notice that we do not 
recover the single-field squeezed limit result of \cite{Maldacena:2002vr}. This is to be expected since 
our $\f$ is the local one produced outside the horizon. As a consequence, we have only used 
the 
first and 
second-order horizon-crossing contributions coming from the redefinitions in the cubic action 
(see \cite{Tzavara:2010ge,Tzavara:2011hn}) 
as initial sources of $\f$ in the context of the 
long-wavelength formalism. 
The result of \cite{Maldacena:2002vr} on the other hand comes from the interaction 
terms in the Langrangian and hence is not the same.} 
 
The decrease of $\f$ for more squeezed triangles seems contradictory 
to the well-known fact that the local bispectrum is maximized 
for squeezed configurations. In order to clarify this subtle point, we stress that 
the left-hand side of figure \ref{fig3} is essentially the ratio of the exact 
bispectrum to the bispectrum 
assuming $\f$ as a constant (\ref{bisp}) and hence the products 
of the power spectrum cancel out. 
We also plot on the right-hand side of figure \ref{fig3} the final value of the 
bispectrum (\ref{bisp}), 
normalised by the value of the bispectrum for equilateral triangles 
with $K=3k_{60}/2$. 
Although $\f$ is maximal 
for equilateral triangles, the bispectrum has the opposite 
behaviour, since it is dominated by the contribution of the products 
of the power spectrum, which leads to an increased bispectrum 
for the more squeezed shape. At the same time though we show 
that there is a small contribution of $\f$ itself, leading to 
smaller values of the bispectrum when compared to a bispectrum 
where $\f$ is assumed to be constant.

\begin{figure}
\begin{tabular}{ll}
\includegraphics[scale=0.8]{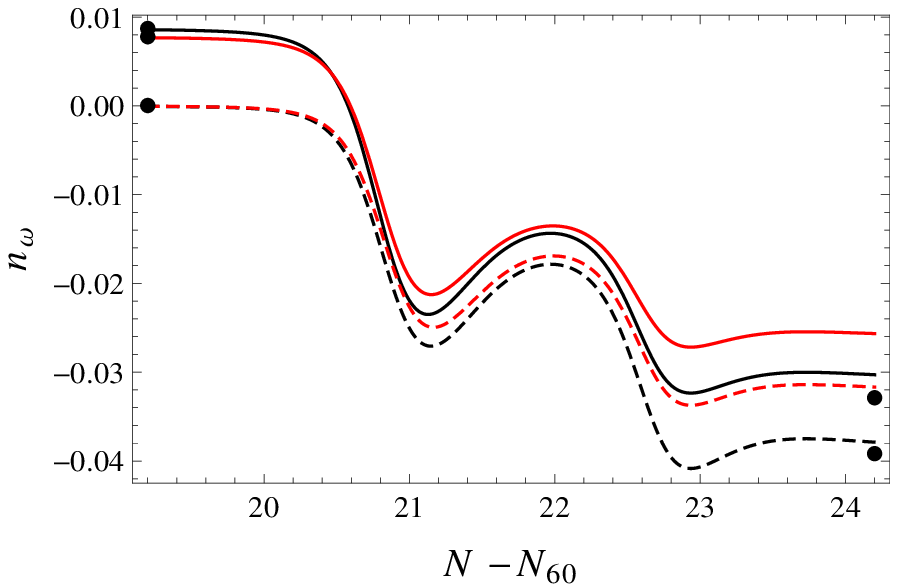}
&\includegraphics[scale=0.8]{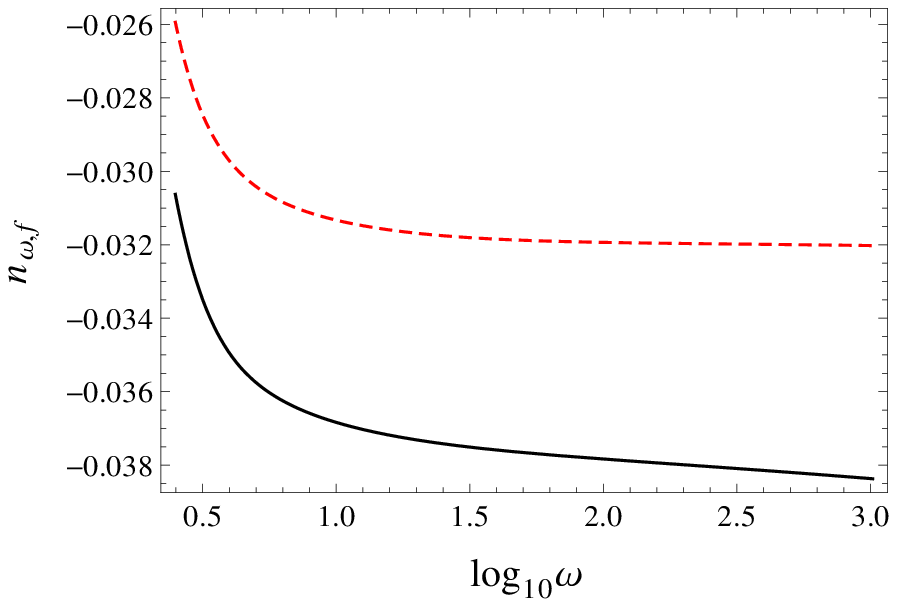}
\end{tabular}
\caption{Left: The time evolution of the shape index 
$n_\omega$ (\ref{intodef}) around the time of the turning of the fields,
for constant $K=(3/2)k_{60}$ (black curves) 
and $K=(3/2)k_{60}/10$ (red curves, above the $K=(3/2)k_{60}$ curves), 
and for the shapes $\omega=5/2$ (solid curve) 
and $\omega=1000$ (dashed curve). 
The points on the left correspond to the analytical values 
of the index as calculated from (\ref{noin}), while the points on the right correspond to the values of the index as calculated from (\ref{A2}) for $\omega\mg 1$.
Right: The final value of the shape index $n_\omega$ 
(\ref{intodef}) for constant $K=(3/2)k_{60}$ (black) and $K=(3/2)k_{60}/10$ (dashed red) 
as a function of $\omega$. Both figures are for the model (\ref{qua}) with mass 
ratio $m_{\phi}/m_{\gs}=9$.}\label{fig41}
\end{figure}

In order to quantify the above results, we examine the shape index $\tn_{\omega}$ (\ref{inddef}), 
assuming $K=\mathrm{const}$ and $\gamma=0$,
\be
 \tn_{\omega}=\frac{\partial\ln{f}}{\partial t_{k'}}\frac{1}{1-\e_{k'}}
 \frac{1}{1+2\omega}
 -\frac{\partial\ln{f}}{\partial t_{k}}\frac{1}{1-\e_{k}}
 \frac{2\omega}{1+2\omega}
.\label{shap}
\ee
In terms of $\tn_{\omega}$, $\f$ takes the form
\be
-\frac{6}{5}\f=\frac{1}{(2\pi^2)^2}\frac{f(k'_0,k'_0)+
2\omega^3f(k'_0,k_0)\lh\frac{\omega}{\omega_0}\rh^{\tn_{\omega}(t_{k'_0},t_{k_0})}}
{\mathcal{P}_\zeta(k'_0)^2\lh\frac{k'}{k'_0}\rh^{2(n_\zeta(t_{k'_0})-1)}
+2\omega^3\mathcal{P}_\zeta(k'_0)\mathcal{P}_\zeta(k_0)\lh\frac{k'}{k'_0}\rh^{n_\zeta(t_{k'_0})-1}\lh\frac{k}{k_0}\rh^{n_\zeta(t_{k_0})-1}},
\label{nbf}
\ee
where $k'/k'_0=(1+\beta)/(1+\beta_0) \propto 2\omega/(\omega+\frac{1}{2})$ and 
$k/k_0=(1-\beta)/(1-\beta_0) \propto 1/(\omega+\frac{1}{2})$. 
This can be further simplified in the squeezed region $\omega\mg1$ to find
\bea
-\frac{6}{5}\f&=&\frac{1}{(2\pi^2)^2}\frac{f(k'_0,k_0)\lh\frac{\omega}{\omega_0}\rh^{\tn_{\omega}(t_{k'_0},t_{k_0})}}
{\mathcal{P}_\zeta(k'_0)\mathcal{P}_\zeta(k_0)
\lh\frac{\omega(\omega_0+\frac{1}{2})}{\omega_0(\omega+\frac{1}{2})}\rh^{n_\zeta(t_{k'_0})-1}
\lh\frac{\omega_0+\frac{1}{2}}{\omega+\frac{1}{2}}\rh^{n_\zeta(t_{k_0})-1}}\nn\\
&\equiv&
-\frac{6}{5}f_{\mathrm{NL},0}
\lh\frac{\omega}{\omega_0}\rh^{\tn_{\omega}(t_{k'_0},t_{k_0})+n_\zeta(t_{k_0})-1}
\equiv
-\frac{6}{5}f_{\mathrm{NL},0}
\lh\frac{\omega}{\omega_0}\rh^{n_{\omega}(t_{k'_0},t_{k_0})},
\label{into}
\eea
with
\be
n_{\omega}\equiv\frac{\d\ln\f}{\d\ln\omega}=\frac{\partial\ln{\f}}{\partial t_{k'}}\frac{1}{1-\e_{k'}}
\frac{1}{1+2\omega}
-\frac{\partial\ln{\f}}{\partial t_{k}}\frac{1}{1-\e_{k}}
\frac{2\omega}{1+2\omega}
.\label{intodef}
\ee
The shape index $n_\omega$ describes the change of $\f$ due to the relative size 
of the two scales, namely due to how squeezed the triangle is, while 
keeping $K$ constant (see the right-hand side of figure 
\ref{fig01}). 

We studied different squeezed  
triangle configurations with constant $K$, varying $\omega$ from $\omega=5/2$ to $\omega=1000$. 
On the left-hand side of figure~\ref{fig41} we plot the time evolution of the shape index. 
The negative values of the index signify the decrease of $\f$ as expected. 
As one can see from the figure, for the more squeezed triangle ($\omega=1000$)
the initial value of $n_\omega$ seems to depend solely on the shape of the 
triangle and not on its magnitude, and even for the less squeezed triangle
($\omega=5/2$) the initial dependence on $K$ is negligible.  
We can find the analytical initial value of $n_\omega$ by differentiating (\ref{fnlin}):
\be
n_{\omega,in}=
\frac{1}{1+2\omega}n_{K,in}+\frac{4\omega}{1+2\omega}G_{22k'k}\frac{(\hpe_{k'})^2}{\e_{k'}+\hpa_{k'}}.\label{noin}
\ee
For $G_{22k'k}=0$, which corresponds to the squeezed limit, $n_{\omega,in}$ is proportional to the initial shape index for equilateral 
triangles times a factor depending on the shape, which also becomes very small in the squeezed limit.

Super-horizon effects, and especially the turning of the fields, 
result in a separation of the curves of $n_\omega$ of the same shape for 
different values of $K$, due to the dependence of the evolution of $\bv_{12k'}$ 
on the scale $k'$. 
The turning of the fields increases the absolute value of $n_\omega$, which is
the opposite of the behaviour of the conformal index $n_K$. 
The shape index depends on the transfer function 
$\bv_{12k'}$ (see Appendix \ref{app} for an analytical approximation). 
The smaller $K$, the less does the final value of $\bv_{12k'}$ change with respect to its initial value (see figure 
\ref{fig1}) and hence the less the shape index is affected.  
Notice that the slow-roll parameters at horizon-crossing 
have the opposite behaviour: the smaller $K$, the smaller they are. Even though $n_\omega$ also depends on the slow-roll parameters, 
it is $\bv_{12k'}$ that most affects its evolution.

On the right-hand side of figure \ref{fig41} we plot 
the value of the shape index at the end of inflation.  
It exhibits a running of about $20\%$ within the 
range of scales studied, somewhat larger than the conformal index. 
We have analytically computed the shape spectral index for models with $\tg_{int}=0$ and with final $g_{iso,f}=0$ 
and give the result in Appendix \ref{app}.
 
The dotted curve in the plot on the left-hand side of figure~\ref{fig3} shows
the final value of $\f$ approximated as a simple power law according to 
(\ref{into}).
Within the range of validity of 
our approximation $\omega^3\mg1$ it describes the exact result very well. 
We have also studied the shape spectral index for the potential 
(\ref{pot}). Similarly to $n_K$, its value is two orders of magnitude smaller than the value for the quadratic potential, 
but the parametrization of 
$\f$ in terms of the shape index is in good agreement with the exact result for a larger range of $\omega\gtrsim3/2$.

\section{Conclusions}\label{concl}

In this paper we studied the scale dependence of the local non-Gaussianity
parameter $\f$ for 
two-field inflationary models. Multiple-field models with standard kinetic terms
do not exhibit 
the strong scale dependence inherent in models that produce equilateral non-Gaussianity 
at horizon-crossing through quantum mechanical effects. Nevertheless they are not 
scale independent in general and the interesting question is whether we can  
profit from their scale dependence in order to observationally acquire more 
information about inflation. 

We have calculated $\f$ using the long-wavelength formalism. This 
constrains us to assume slow roll at horizon-crossing and hence the relevant 
quantities at that time should not vary much, including the 
scale dependence of $\f$ for any triangle shape. Indeed we confirmed that, 
by introducing the conformal spectral index $n_K$ that measures the tilt of $\f$
for triangles of the same shape but different size ($K$ is a variable 
proportional to the perimeter of the momentum triangle). For the 
quadratic model with mass ratio $m_{\phi}/m_{\gs}=9$ we find $n_K\simeq0.018$, 
pointing to an almost scale-invariant $\f$. 

We also studied the scale dependence of $\f$ while varying the shape of the 
triangle and keeping its perimeter constant. $\f$ exhibits the opposite behaviour of the full bispectrum, 
i.e.\ it decreases the more squeezed the triangle is (the momentum dependence
of the bispectrum is dominated by that of the products of power spectra, not by
that of $\f$). This variation 
is not related to horizon-crossing quantities, but rather to the fact that 
the more squeezed the isosceles triangle under study, 
the smaller the correlation of its two scales. 
We quantified this effect by introducing the shape spectral index $n_\omega$, which 
for the quadratic model with $m_{\phi}/m_{\gs}=9$ is $n_{\omega}\simeq-0.03$ and has a running of about 
$20\%$ ($\omega$ is defined as the ratio of the
two different sides of an isosceles momentum triangle).

All our calculations have been done numerically in the exact background, 
assuming slow roll only at horizon-crossing, not afterwards. 
Nevertheless, semi-analytical expressions can be easily produced by directly 
differentiating $\f$. If we do assume slow roll we showed that we can even simplify these expressions further 
and find analytical 
formulas for the final value of $\f$ and its spectral indices $n_K$ and 
$n_\omega$, if  
the integral in $\f$ and the isocurvature modes vanish by the end of inflation,
which is the case for example for any equal-power sum model. 

We used the two-field quadratic potential in our numerical calculations.
This potential is easy to examine and allows for simplifications in the relevant 
expressions. Although its final non-Gaussianity is small, $\mathcal{O}(\e_k)$, its 
general behaviour should not be different from other multiple-field 
inflationary models with standard kinetic terms, in the sense 
that the scale dependence of $\f$ 
should always depend on horizon-exit quantities and the evolution of the transfer functions during the turning of the fields.  
Indeed we have checked that for the potential (\ref{pot}) studied in 
\cite{Tzavara:2010ge}, able to produce $\f\sim\mathcal{O}(1)$, the results 
remain qualitatively the same, although the values of the spectral indices are smaller due to the very slow evolution 
of the background at horizon-crossing in that model.

Although 
the effect of the magnitude of the triangle on $\f$ had been considered before, analytical 
and numerical estimates were not available before this paper. In addition, it is the first time that 
the dependence of $\f$ itself (instead of the power spectra in the bispectrum) on the shape of the momentum triangle is studied. Using the long-wavelength 
formalism we have managed to study the two different sources of momentum dependence, i.e.\ the slow-roll parameters at horizon 
crossing and the evolution of the transfer functions, and to understand the role of each for the two different triangle deformations 
that we have studied. In summary, the later a momentum mode exits the horizon, the larger the slow-roll parameters are at that time and 
the larger $\f$ tends to be. In contrast, the final value of $\bv_{12k'}$ and the initial value of $G_{22k'k}$, the 
two transfer functions that are the most important for $\f$, are smaller the later the scale exits, which results 
in decreasing values of $\f$. These two opposite effects manifest themselves in the two different deformations we have studied. When keeping the shape of 
the triangle constant and varying its size, it is the slow-roll parameters at horizon crossing that play the major role in $\f$ and result in an increasing $\f$ for  
larger $K$. When changing the shape of $\f$, it is the correlation between the isocurvature mode at different scales, $G_{22k'k}$, that has the most 
important role, resulting in decreasing values of $\f$ when squeezing the triangle (i.e.\ increasing $\omega$). 

We have verified that the spectral indices of $\f$ ($n_K$ and
$n_\omega$), which we introduced to describe the effect of the two
types of deformations of the momentum triangle, provide a good
approximation over a wide range of values of the relevant scales.  In
the models we studied their values are too small to be detected by
Planck, given that $\f$ itself cannot be big (or it would have been
detected by Planck).  Models that break slow roll at horizon crossing
could in principle have larger spectral indices, but in order to study
such models one would need to go beyond the long-wavelength
formalism. Such models could be studied using the
exact cubic action derived in \cite{Tzavara:2011hn}.

\appendix

\section{Analytical expressions for the spectral indices}
\label{app}

By differentiating (\ref{fnlfin}) and using (\ref{con}) we can find the final value of $n_K$ for equilateral triangles in the slow-roll approximation, 
assuming that isocurvature modes have vanished 
for an equal-power sum potential (for which the $\tg_{int}$ contribution is zero, see (\ref{tgint}) and \cite{Tzavara:2010ge}):
\bea
&&n_{K,eq,f}=\nn\\
&&-4\frac{\bv_{12k}(\bv_{12k}\chi_k-2 \hpe_k)}{1+(\bv_{12k})^2}
-\frac{1}{f_{\mathrm{NL},eq,f}(1+(\bv_{12k})^2)^2} \Bigg[-2\e_k^2-3\e_k\hpa_k+(\hpa_k)^2+5(\hpe_k)^2-\xi^\parallel_k\nn\\
&&
+3\bv_{12k}\Bigg(\hpe_k(3\e_k+6\hpa_k-5\chi_k)-\xi^\perp_k\Bigg)+3(\bv_{12k})^2\Big(\tilde{W}_{221k}+4(\hpe_k)^2
-2(\e_k+\hpa_k-\chi_k)(\e_k+2\chi_k)\Big)\nn\\
&&
+\frac{(\bv_{12k})^3}{\hpe_k} 
\Bigg(\chi_k\Big(3\e_k^2-2\tilde{W}_{221k}+4 \e_k\hpa_k +3 (\hpa_k)^2 - 8(\hpe_k)^2 +\hpa\chi_k -3\chi_k^2\Big)
+\xi^\parallel_k(\e_{k}+\hpa_{k}-\chi_{k})\nn\\
&&+\hpa_k(\e_k^2 - (\hpa_k)^2)+\tilde{W}_{221k}(\e_k+\hpa_k)+(\hpe_k)^2(2\e_k+5\hpa_k)-\xi^\perp_k\lh\hpe_{k}+\frac{(\e_{k}+\hpa_{k}+\chi_{k})\chi_{k}}{\hpe_{k}}\rh
\Bigg)
\Bigg],\label{A1}
\eea
where $\tilde{W}_{221}=(\sqrt{2\e}/\kappa)W_{221}/(3H^2)$. 
We have checked this approximation and we find good agreement with the exact conformal index for equilateral triangles.

We repeat the calculation for the shape index $n_\omega$ (\ref{intodef}), differentiating the squeezed $\f$ (\ref{fnlb}). Where needed we use the slow-roll 
approximation $G_{32k'k}=-\chi_{k'}G_{22k'k}$ and $G_{23k'k}=G_{22k'k}/3$.
The result is:
\bea
 &&n_{\omega,sq,f}=\label{A2}\\
 &&\frac{1}{f_{\mathrm{NL},sq,f}(1+(\bv_{12})^2)^2}\frac{1}{1+2\omega}\Bigg\{
 \frac{2G_{22k'k}\bv_{12k'}}{1+(\bv_{12})^2}\lh4-\lh\omega
 +(2+\omega)\bv_{12k'}\rh\frac{\chi_{k'}}{\hpe_{k'}}\rh\nn\\
 &&\times\Bigg[\hpe_{k'}+\bv_{12k'}\lh3(\e_{k'}+\hpa_{k'})-2\chi_{k'}\rh+(\bv_{12k'})^2\lh \hpe_{k'}-\frac{(\e_{k'}+\hpa_{k'}-\chi_{k'})
 \chi_{k'}}{\hpe_{k'}}\rh\Bigg]
 \nn\\
 &&-G_{22k'k}\Bigg[2(\hpe_{k'})^2+\bv_{12k'}\lh-\xi^\perp_{k'}+\hpe_{k'}(11\e_{k'}+14\hpa_{k'}-8\chi_{k'})\rh
 +\frac{(\bv_{12k'})^3}{\hpe_{k'}}\Bigg(
 \chi_{k'}\Big(3\e_{k'}^2-2\tilde{W}_{221k'}\nn\\
 &&
 +4\e_{k'}\hpa_{k'}+3(\hpa_{k'})^2-7(\hpe_{k'})^2
 -2\chi_{k'}^2-\e_{k'}\chi_{k'}\Big)+\xi^\parallel(\e_{k'}+\hpa_{k'}-\chi_{k'})+\hpa_{k'}(\e_{k'}^2-(\hpa_{k'})^2)\nn\\
&&+\tilde{W}_{221k'}(\e_{k'}+\hpa_{k'})
 +(\hpe_{k'})^2(2\e_{k'}+5\hpa_{k'})
 -\xi^\perp_{k'}\lh\hpe_{k'}+\frac{(\e_{k'}+\hpa_{k'}+\chi_{k'})\chi_{k'}}{\hpe_{k'}}\rh \Bigg)\nn\\
 &&+(\bv_{12k'})^2\lh2\tilde{W}_{221k'}-6\e_{k'}^2+(\hpa_{k'})^2+9(\hpe_{k'})^2-\xi_{k'}^\parallel-9\hpa_{k'}\chi_{k'}+8\chi_{k'}^2-\e_{k'}(7\hpa_{k'}+5\chi_{k'})\rh
 \Bigg]\nn\\
 &&
 -\frac{1}{1+(\bv_{12k'})^2}\Bigg[(\hpa_{k'})^2+3(\hpe_{k'})^2-2\e_{k'}^2-3\e_{k'}\hpa_{k'}-\xi_{k'}^\parallel
 -2(\bv_{12k'})^3\lh\xi_{k'}^\perp+\hpe_{k'}(\e_{k'}-2\hpa_{k'}-5\chi_{k'})\rh\nn\\
 &&
 +(\bv_{12k'})^4\lh \tilde{W}_{221k'} +(\e_{k'}-\hpa_{k'})\hpa_{k'}+3(\hpe_{k'})^2+\xi_{k'}^\parallel+2\chi_{k'}(\e_{k'}+\chi_{k'})\rh
 -2\bv_{12k'}\big( \xi_{k'}^\perp+\hpe_{k'}(5\e_{k'}\nn\\
 &&+2\hpa_{k'}+3\chi_{k'})\big)
 +(\bv_{12k'})^2\lh \tilde{W}_{221k'} +2\lh-\e_{k'}(\e_{k'}+\hpa_{k'})-5(\hpe_{k'})^2+(3\e_{k'}+3\hpa_{k'}+\chi_{k'})\chi_{k'}\rh\rh
 \Bigg]\Bigg\}\nn
 \eea
We have checked this approximation and we find good agreement with the exact shape index for $\omega\gtrsim3$.

\bibliography{bib}{}

\bibliographystyle{utphys.bst}

\end{document}